\RequirePackage[2020-02-02]{latexrelease}
\documentclass[aps,prd,superscriptaddress,showpacs,preprint]{revtex4}
\usepackage{graphicx, bm}
\usepackage[usenames]{color}
\usepackage{float}
\usepackage{multirow}
\usepackage{subcaption}

\usepackage{slashed}

\begin{document}

%\tightenlines
\draft

\title{Search for pseudoscalar Higgs boson $A_0$ of the Bestest Little Higgs model at the LHC and FCC-hh }

\author{Y. Cervantes-Baltazar\footnote{yahiralejandro.708805@fisica.uaz.edu.mx}}
\affiliation{\small Facultad de F\'{\i}sica, Universidad Aut\'onoma de Zacatecas\\
         Apartado Postal C-580, 98060 Zacatecas, M\'exico.\\}

\author{E. Cruz-Albaro\footnote{elicruzalbaro88@gmail.com}}
\affiliation{\small Facultad de F\'{\i}sica, Universidad Aut\'onoma de Zacatecas\\
            Apartado Postal C-580, 98060 Zacatecas, M\'exico.\\}

\author{ A. Guti\'errez-Rodr\'{\i}guez\footnote{alexgu@fisica.uaz.edu.mx}}
\affiliation{\small Facultad de F\'{\i}sica, Universidad Aut\'onoma de Zacatecas\\
         Apartado Postal C-580, 98060 Zacatecas, M\'exico.\\}

\author{ M. A. Hern\'andez-Ru\'{\i}z\footnote{mahernan@uaz.edu.mx}}
\affiliation{\small Unidad Acad\'emica de Ciencias Qu\'{\i}micas, Universidad Aut\'onoma de Zacatecas\\
         Apartado Postal C-585, 98060 Zacatecas, M\'exico.\\}

\author{D. Espinosa-G\'omez \footnote{david.espinosa@umich.mx}
}
\affiliation{\small Facultad de Ciencias F\'{\i}sico Matem\'aticas, Universidad Michoacana de San Nicol\'as de Hidalgo\\
            Avenida Francisco, J. M\'ujica S/N, 58060, Morelia, Michoac\'an, M\'exico.\\}

\date{\today}
%\maketitle
%\RequirePackage[2020-02-02]{latexrelease}

\begin{abstract}
% Insert abstract here

We analyse the sensitivity of the Large Hadron Collider (LHC) and the Future Circular Collider-hadron-hadron (FCC-hh) to the existence
of pseudoscalar Higgs boson $A_0$ predicted by the Bestest Little Higgs model. We study the tree-level (two and three body) and one-loop decays of the
pseudoscalar, $A_0 \to b\bar b, t \bar t, \gamma t\bar t, \gamma b \bar b, Zt\bar t, Zb \bar b, Wt\bar b, h_0 t \bar t, h_0 b \bar b$, and
 $A_0 \to \gamma\gamma, \gamma Z, ZZ, gg, WW$, respectively. In addition, we perform a phenomenological study of the production of the pseudoscalar $A_0$ via gluon fusion $gg \to A_0 \to Y$ processes, where $Y \equiv \gamma\gamma, \gamma Z, ZZ, gg, WW$. From the cross section of the processes of interest and the expected integrated luminosity of the LHC and  FCC-hh, we determined
the number of events that could be produced in both colliders.   \\

\end{abstract}


\vspace{5mm}

\maketitle

%\narrowtext

\section{Introduction}

The discovery of a Higgs boson with a mass of 125 GeV by the ATLAS~\cite{ATLAS:2012yve}  and CMS~\cite{CMS:2012qbp} collaborations at the CERN LHC has motivated the search for new exotic states at current and future colliders.
Another significant fact is the recent observation of the production of four top quarks in proton-proton collisions at a center-of-mass energy of 13 TeV~\cite{CMS:2023ftu,ATLAS:2023ajo}. This is one of the rarest processes in the Standard Model (SM) that is currently accessible at hadron colliders.  The top quark, the heaviest elementary particle in the SM, has strong connections to the SM Higgs boson and new particles predicted in theories beyond the SM (BSM). Many of these new heavy particles can decay into top quarks, resulting in the production of four top quarks via intermediate exotic particles. Such a signature could indicate the existence of new BSM particles.

The unsolved puzzles of the SM (such as dark energy, dark matter, neutrino masses, and the matter-antimatter asymmetry, among others)  require new BSM physics. Many of these theories necessitate a richer Higgs sector. This extended sector is a common feature of several extended models. The Higgs sector itself has become a fundamental tool in the search for new physics, which could manifest in various ways. For example, the discovery of additional Higgs bosons or the measurement of new sources of CP violation in the scalar sector would provide direct evidence of BSM physics~\cite{Englert:2014uua}.
So far, both direct and indirect searches for new physics have produced unsuccessful results. However, it is expected that the scope of these searches will expand considerably in the future, as colliders will operate at such high energy scales that new signal channels will open up which would be suppressed at lower energies. The discovery of new non-SM Higgs bosons would be a clear sign of BSM physics with extended sectors. 

Several extended models have been proposed to address open issues in the SM scenario, such as the hierarchy problem~\cite{Schmaltz:2002wx}. This problem is also known as the fine-tuning problem. Among the BSM models, there are Little Higgs-type models~\cite{Arkani-Hamed:2002ikv,Chang:2003zn,Han:2003wu,Chang:2003un,Schmaltz:2004de} that introduce enough new physics to generate cancellations and preserve a light Higgs boson.  These models implement the idea that the Higgs boson is a pseudo-Goldstone boson. Little Higgs  theories are constructed by embedding the SM within a larger group with an extended symmetry.  From a phenomenological point of view, this symmetry requires the existence of new particles whose couplings ensure that large contributions to the Higgs mass cancel out. Thus, the Higgs boson mass is protected from radiative corrections.
Although Little Higgs models offer an attractive solution to the hierarchy problem, many of them exhibit certain theoretical inconsistencies~\cite{Schmaltz:2008vd}. For example, the mechanism used to generate a Higgs quartic coupling leads to violations of custodial symmetry. They are also strongly constrained by electroweak precision data in the gauge sector.

The Bestest Little Higgs model (BLHM)~\cite{JHEP09-2010,Godfrey:2012tf,Kalyniak:2013eva} was formulated to resolve the theoretical inconsistencies found in most Little Higgs theories. This model introduces separate symmetry-breaking scales, $f$ and $F$, where $F > f$. In this way, the new heavy quarks ($ T $,
 $ T_5 $, $ T_6 $, $ T^{2/3} $, $ T^{5/3} $,  $ B $) acquire masses proportional to the $f$ scale, while the new gauge bosons ($ Z' $, $ W'^{\pm} $) acquire masses proportional to a combination of the $f$ and $F$ scales. These gauge boson partners allow the model to evade the constraints imposed by electroweak precision measurements. A successful quartic Higgs coupling is also generated without introducing any dangerous singlet in the scalar sector~\cite{Schmaltz:2008vd}, in contrast to other Little Higgs models. Instead, an additional singlet (satisfying certain additional symmetries) and a custodial $SU(2)$ symmetry~\cite{JHEP09-2010}.
The scalar sector of the BLHM has a rich phenomenology that generates neutral and charged scalar fields: $h_0, H_0, A_0, \phi^{0},\eta^{0}, H^{\pm}, \phi ^{\pm}$, and $ \eta^{\pm}$. 
We recommend that interested readers refer to Refs.~\cite{Martinez-Martinez:2024lez,Cruz-Albaro:2024vjk,Cruz-Albaro:2022lks} for more information on the BLHM.

In this work, we investigate the production of the pseudoscalar Higgs boson $A_0$ via gluon fusion at current and future colliders, such as the LHC~\cite{ZurbanoFernandez:2020cco,Azzi:2019yne,Cepeda:2019klc,FCC:2018bvk} and the FCC-hh~\cite{MammenAbraham:2024gun,Mangano:2022ukr,FCC:2018byv,FCC:2018vvp}.  
Specifically, we discuss searching for pseudoscalar $A_0$ decays into gauge bosons ($V V$, where $V = \gamma, Z, W, g$) within the context of the BLHM.
 Although $A_0 \to VV$ decay is an induced process at  one-loop level, we believe the  pseudoscalar  could be first detected in a two-photon final state, similar to the discovery of the SM Higgs boson~\cite{ATLAS:2012ad,CMS:2012qbp,ATLAS:2015yey}.
Searches in the di-photon channel are especially useful as it offers the best mass resolution and an exceptionally clean signal because the CMS and ATLAS detectors reconstruct photons with high precision.
 However, even if the pseudoscalar Higgs $A_0$ is discovered in other final states, the other possible decays must be confirmed or excluded. Thus, this study will provide information about the BLHM  and the potential detection of the pseudoscalar $A_0$ at the LHC and the future 100 TeV collider. 
 
 The physics programs of current and future hadron, lepton, and hybrid colliders include, among their research objectives, the theoretical, phenomenological, and experimental study of exotic particles, such as the pseudoscalar $A_0$ predicted by several extensions of the SM. One of these extensions is the BLHM. For further information on studies of the pseudoscalar within other BSM frameworks, we recommend that interested readers consult Refs.~\cite{Arhrib:2018pdi,Abu-Ajamieh:2025zcv,Aranda:2017bgq}. 
 On the other hand, experimental searches for pseudoscalar Higgs bosons are essential, as they provide complementary information to theoretical predictions and help constrain the parameter space of BSM models~\cite{Cornell:2020usb,Almarashi:2021pgu}. Some searches for the pseudoscalar have been conducted in proton-proton collisions at center-of-mass energies of 8 and 13 TeV with the ATLAS detector, where the pseudoscalar $A_0$ is produced in association with a top quark pair, followed by the decay $A_0\to t\bar{t}$ (i.e., $t\bar{t}A_0 \to t\bar{t}t\bar{t}$)~\cite{ATLAS:2024jja,ATLAS:2022rws,ATLAS:2017snw}.
Another possible production mechanism for the pseudoscalar occurs in association with a $Z$ gauge boson via gluon fusion at the LHC ($gg \to Z A_0 \to l \bar{l} b\bar{b}+X$)~\cite{Kao:2003jw}.
 The results of these search channels have been interpreted within the framework of the Two-Higgs Doublet model (2HDM); however,  they can also be investigated within the BLHM scenario. These studies are currently underway, and we expect to publish the results soon.
These different discovery channels offer valuable opportunities to search for the pseudoscalar both at the LHC and at future colliders and may provide new insights into BSM physics.

The article is organized as follows.  In Section~\ref{A0decay}, we present the analytical calculations of the tree-level and one-loop decays of the pseudoscalar Higgs boson $A_0$. In Section~\ref{results}, the numerical results are discussed. Finally, conclusions are presented in Section~\ref{conclusions}. The  effective couplings involved in our calculations are provided in  Appendix~\ref{rulesF}.

\section{ Pseudoscalar  $A_0$ decays in the BLHM} \label{A0decay}

In this section, we determine the partial decay widths of the pseudoscalar $A_0$, or otherwise provide the corresponding transition amplitudes. In this regard, we will consider direct search channels with final states of the SM. In the following, we describe the different decay modes of the Higgs boson pseudoscalar $A_0$.

\subsection{Two-body decays  at tree level}

\begin{figure}[H]
\begin{center}
 \includegraphics[width=4.0cm,height=2.0cm]{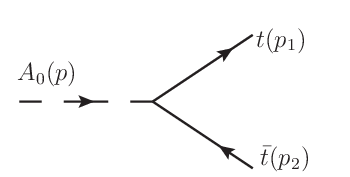}
  \includegraphics[width=4.0cm,height=2.0cm]{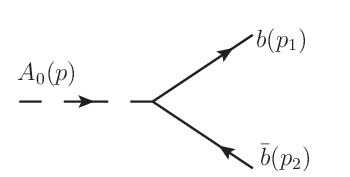}
\caption{\label{tree1} Feynman diagrams corresponding to the tree-level two-body decays of the  pseudoscalar $A_0$.}
\end{center}
\end{figure}

The Feynman diagrams corresponding to the tree-level two-body decays of the pseudoscalar $A_0$ are shown in Fig.~\ref{tree1}. To obtain the analytical expressions for the decay widths, we use the Feynman rules for the interaction vertices as given in Refs.~\cite{Cruz-Albaro:2022lks,Cruz-Albaro:2023pah,Cruz-Albaro:2022kty,Gutierrez-Rodriguez:2023sxg}, with the corresponding effective couplings detailed in Appendix~\ref{rulesF}. Next, in Eqs.~(\ref{dw1}) and (\ref{dw2}) we provide the resulting expressions for the partial decay widths

\begin{eqnarray}
  \Gamma(A_0 \to t \bar{t}) &=&  \frac{ N_C\, g_{A_0 tt}^2 m_{A_0}}{8 \pi  }  \left(1-\frac{4  m_{t}^{2}}{m_{A_0}^{2}} \right)^{1/2}, \label{dw1} \\ 
  \Gamma(A_0 \to b \bar{b}) &=& \frac{ N_C \, g_{A_0 bb}^2 m_{A_0} }{8 \pi}  \left(1-\frac{4  m_{b}^{2}}{m_{A_0}^{2}} \right)^{1/2}. \label{dw2}
\end{eqnarray}

\noindent The colour factor of the quarks is represented by $ N_C$  in these expressions, while the effective couplings are denoted by $g_{A_0 tt}$ and $g_{A_0 bb}$ (see Appendix~\ref{rulesF}).

\subsection{Three-body decays  at tree level}

\begin{figure}[H]
\begin{center}
  \includegraphics[width=3.5cm,height=2.5cm]{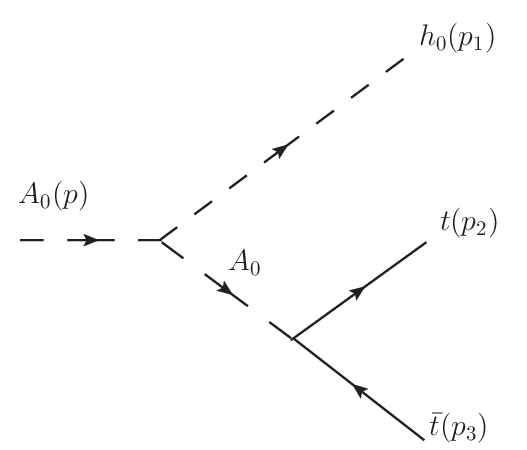}
  \includegraphics[width=3.5cm,height=2.5cm]{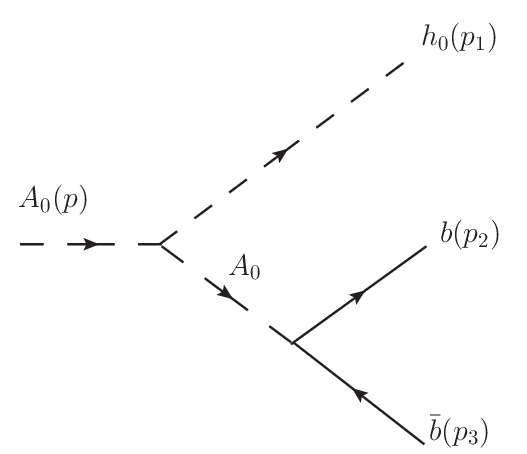}
  \includegraphics[width=3.5cm,height=2.5cm]{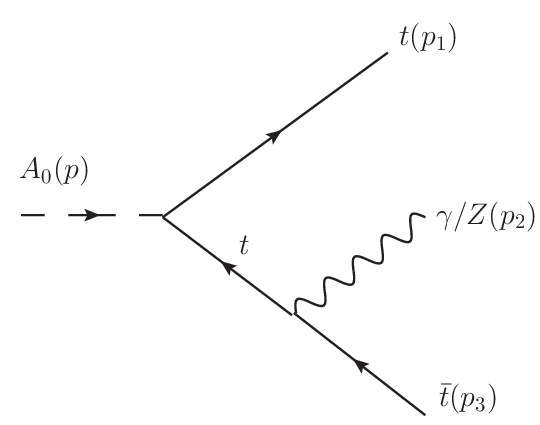}
  \includegraphics[width=3.5cm,height=2.5cm]{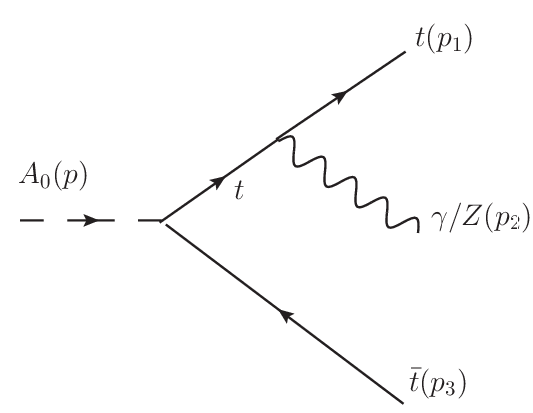}\\
 \includegraphics[width=3.5cm,height=2.5cm]{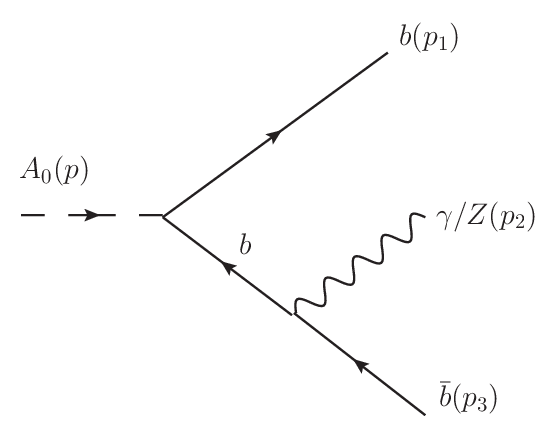}
  \includegraphics[width=3.5cm,height=2.5cm]{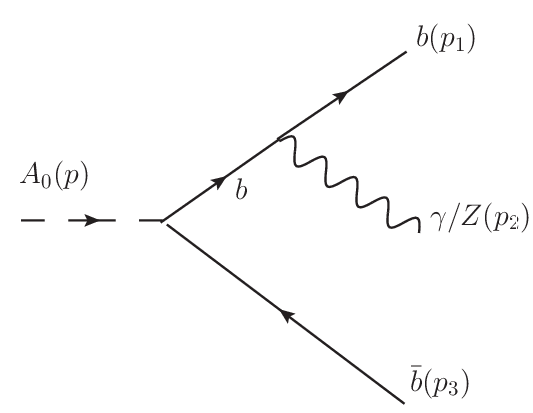}
  \includegraphics[width=3.5cm,height=2.5cm]{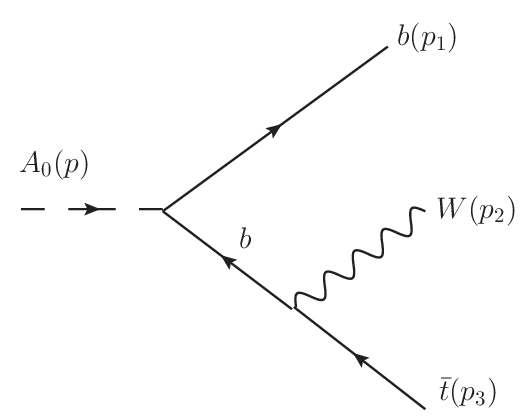}
   \includegraphics[width=3.5cm,height=2.5cm]{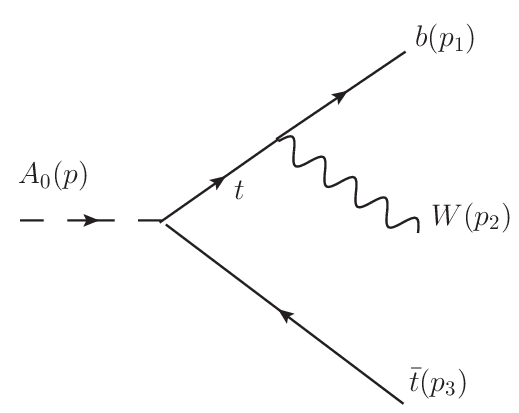}
\caption{\label{tree2} Feynman diagrams corresponding to the tree-level three-body decays of the  pseudoscalar $A_0$.}
\end{center}
\end{figure}

Tree-level three-body decays are fundamental processes in particle physics, as they correspond to the leading-order contributions in the perturbative expansion and can yield significant effects in various scenarios. In the context of the BLHM, the Feynman diagrams corresponding to the three-body decays of the pseudoscalar $A_0$ are shown in Fig.~\ref{tree2}. For these processes, we provide only the decay amplitudes, since the analytical expressions for the decay widths are typically very lengthy. Moreover, the associated phase space is more complex, involving non-trivial multidimensional integrations.
The analytical expressions for the decay amplitudes are calculated using the generic formula described in  Eq.~(\ref{width3})~\cite{pdg:2023,Barger:1996,Barradas:1996xb}:

\begin{eqnarray} \label{width3}
\frac{d\Gamma(A_0 \to ABC)}{dx_a dx_b}=\frac{m_{A_0}}{256\, \pi^3}\vert \mathcal{M}(A_0 \to ABC) \vert^2.
\end{eqnarray}

From the Feynman diagrams (see  Fig.~\ref{tree2}), we derive the following generic amplitudes:

\begin{eqnarray} \label{h0qq}
\mathcal{M}(A_0 \to h_0 q_i \bar{q}_i)=& g_{A_0 A_0 h_0} g_{A_0 q_i q_i} \left(\frac{i}{k^2-m^{2}_{A_0}} \right) \bar{u}(p_2)\gamma^{5} v(p_3),
\end{eqnarray}

\begin{eqnarray} \label{fqq}
\mathcal{M}(A_0 \to \gamma q_i \bar{q}_i) &=& g_{A_0 q_i q_i} g_{\gamma q_i q_i} \bigg[ \bar{u}(p_1)\gamma^{5} \left(\frac{i(-\slashed{k}+m_{q_i})}{k^2-m^{2}_{q_i}} \right) \gamma^{\mu} v(p_3)\nonumber \\
 &+& \bar{u}(p_1)\gamma^{\mu} \left(\frac{i(\slashed{k}+m_{q_i})}{k^2-m^{2}_{q_i}} \right) \gamma^{5} v(p_3) \bigg] \epsilon_\mu^*(p_2),
\end{eqnarray}

\begin{eqnarray} \label{Zqq}
\mathcal{M}(A_0 \to Z q_i \bar{q}_i) &=& g_{A_0 q_i q_i} g_{Z q_i q_i} \bigg[ \bar{u}(p_1)\gamma^{5} \left(\frac{i(-\slashed{k}+m_{q_i})}{k^2-m^{2}_{q_i}} \right) \gamma^{\mu} \left(g^{Z q_i q_i}_{V} + g^{Zq_i q_i}_{A} \gamma^{5}  \right) v(p_3)\nonumber \\
 &+& \bar{u}(p_1) \gamma^{\mu} \left(g^{Zq_i q_i}_{V} + g^{Zq_i q_i}_{A} \gamma^{5}  \right) \left(\frac{i(\slashed{k}+m_{q_i})}{k^2-m^{2}_{q_i}} \right) \gamma^{5} v(p_3) \bigg] \epsilon_\mu^*(p_2),
\end{eqnarray}

\begin{eqnarray} \label{Wbt}
\mathcal{M}(A_0 \to W b \bar{t}) &=& g_{A_0 bb} g_{W b t} \bigg[ \bar{u}(p_1)\gamma^{5} \left(\frac{i(-\slashed{k}+m_b)}{k^2-m^{2}_{b}} \right) \gamma^{\mu} P_L\,  v(p_3)\nonumber \\
 &+& \bar{u}(p_1) \gamma^{\mu} P_L  \left(\frac{i(\slashed{k}+m_t)}{k^2-m^{2}_{t}} \right) \gamma^{5} v(p_3) \bigg] \epsilon_\mu^*(p_2).
\end{eqnarray}

\noindent In these equations, $q_i$ denotes the top and bottom quarks, while $g^{Zq_i q_i}_{V}$ and $g^{Zq_i q_i}_{A}$ represent the vector and axial-vector coupling constants of the $Z$ boson with the $q_i$ quarks, respectively.
The effective couplings involved in Eqs.~(\ref{h0qq})-(\ref{Wbt}) together with the vector and axial-vector couplings, are provided explicitly in Appendix~\ref{rulesF}.

\subsection{Two-body decays  at one-loop level}

\begin{figure}[H]
\begin{center}
 \includegraphics[width=7.5cm,height=2.0cm]{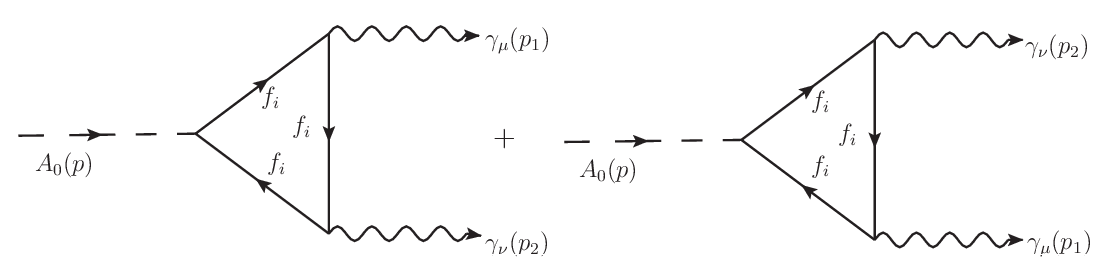}
    \includegraphics[width=7.5cm,height=2.0cm]{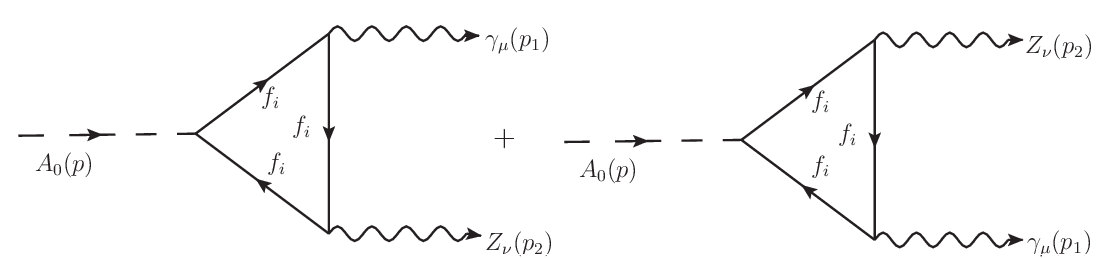}\\
    \includegraphics[width=7.5cm,height=2.0cm]{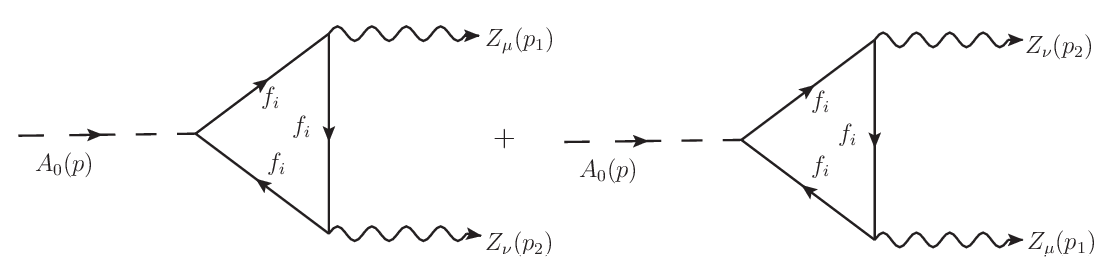}
    \includegraphics[width=7.5cm,height=2.0cm]{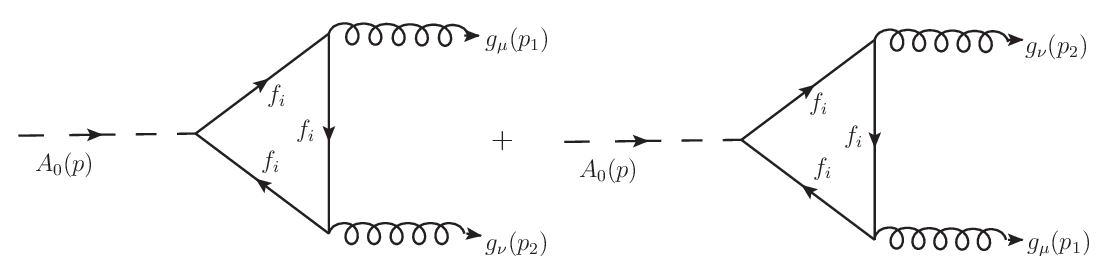}
    \includegraphics[width=7.5cm,height=2.0cm]{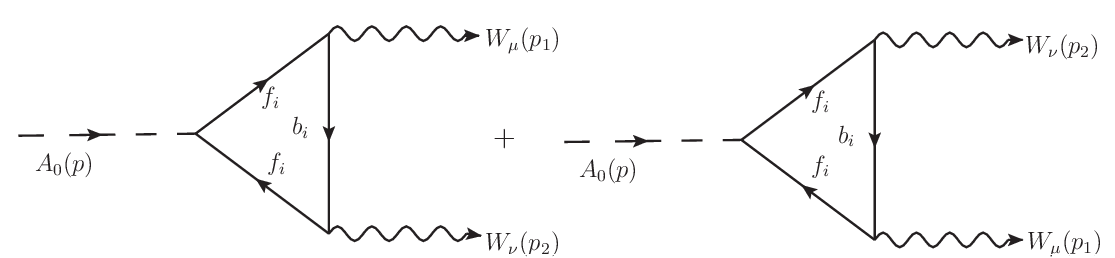}
    \includegraphics[width=7.5cm,height=2.0cm]{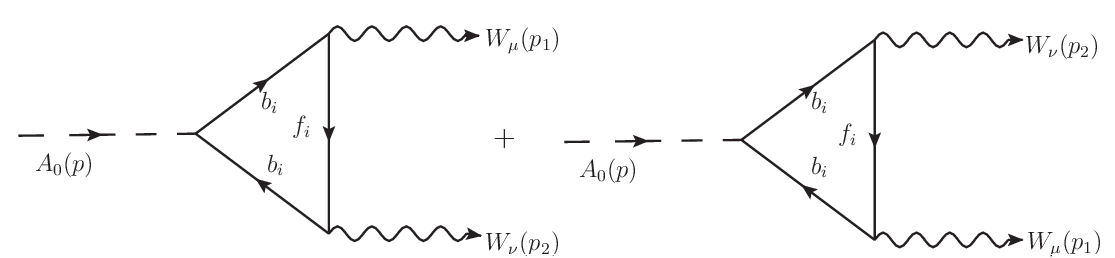}
\caption{\label{loop} Feynman diagrams corresponding to the two-body decays of the  pseudoscalar $A_0$ at one-loop level where $f_i \equiv b, t, T, T_5$, $ T_6 $, $T^{2/3}$, and $b_i \equiv b, B$.}
\end{center}
\end{figure}

Radiative corrections at one-loop level are important processes in particle physics. They allow us to study quantum effects not observed at tree level and make more accurate predictions. Thus, for the pseudoscalar of interest,  we determine the contributions of the partial decay widths of the
 $A_0 \to \gamma \gamma, \gamma Z, ZZ, gg, WW $ processes that arise at one-loop level (see Eqs.~(\ref{wff})-(\ref{w2W})). Fig.~\ref{loop} shows the Feynman diagrams associated with the
  $A_0 \to \gamma \gamma, \gamma Z, ZZ, gg, WW $ decays that are mediated by SM and BLHM fermions: $f_i \equiv b, t, T, T_5$, $ T_6 $, $T^{2/3}$, and $b_i \equiv b, B$.
  The quantum corrections at  one-loop level were computed using the Passarino–Veltman reduction scheme~\cite{Denner:2005nn}. This scheme is a fundamental tool for evaluating loop integrals, as it reduces tensor integrals to scalar integrals, known as the scalar Passarino–Veltman functions. The resulting scalar functions are evaluated using Package-X~\cite{Patel:2015tea,Patel:2016fam}.
  As for the $A_0 \to \gamma \gamma, \gamma Z, ZZ, gg, WW $ decays arising at one-loop level, we have verified that the Feynman diagram contributions are free of ultraviolet divergences.

We find that the generic decay widths of the $A_0 \to \gamma \gamma, \gamma Z, ZZ, gg, WW $ processes at one-loop level are as follows:

\begin{eqnarray} \label{wff}
\Gamma(A_0\to \gamma \gamma) &=&  \frac{m^{3}_{A_0}}{256 \pi ^5}  \bigg| \sum_{i} m_{f_i}\: g_{A_0 f_i f_i}\, g^{2}_{\gamma f_i f_i} C_0(0,0,m_{A_0}^2,m_{f_i}^2,m_{f_i}^2,m_{f_i}^2) \bigg|^{2},
\end{eqnarray}

\begin{eqnarray}
\Gamma(A_0\to \gamma Z) &=& \frac{m^{3}_{A_0} }{128 \pi^5} \left(1-\frac{m^{2}_{Z}}{m^{2}_{A_0}} \right)^{3}  \bigg| \sum_{i} m_{f_i}\: g_{A_0 f_i f_i}\, g_{\gamma f_i f_i}\,  g^{Z f_i f_i}_{V} \nonumber \\
&\times & C_0(m_{A_0}^2,m^{2}_{Z},0, m_{f_i}^2,m_{f_i}^2,m_{f_i}^2)  \bigg|^{2},
\end{eqnarray}

\begin{eqnarray}
\Gamma(A_0\to Z Z) &=&  \frac{1 }{256 \pi^5 m_{A_0}} \left(1-\frac{4 m^{2}_{Z}}{m^{2}_{A_0}} \right)^{-1/2} \bigg| \sum_{i} m_{f_i}\, g_{A_0 f_i f_i} \nonumber \\
&\times & \bigg[  \left(  m^{2}_{A_0} \left((g^{Z f_i f_i}_{A})^{2}-(g^{Z f_i f_i}_{V})^{2} \right)  +4 m^{2}_{Z} (g^{Z f_i f_i}_{V})^{2}\right) C_0(m^{2}_{A_0},m^{2}_{Z},m^{2}_{Z},m^{2}_{f_i},m^{2}_{f_i},m^{2}_{f_i}) \nonumber \\
&+& 4 (g^{Z f_i f_i}_{A})^{2} \left(B_0(m_{A_0}^2, m_{f_i}^2,m_{f_i}^2) -  B_0(m_{Z}^2, m_{f_i}^2,m_{f_i}^2) \right) \bigg] \bigg|^{2},
\end{eqnarray}

\begin{eqnarray}
\Gamma(A_0\to gg) &=& \frac{g_{s}^{4} m^{3}_{A_0}}{128 \pi^5}   \bigg| \sum_{i} m_{f_i}\, g_{A_0 f_i f_i}  C_0(0,0,m^{2}_{A_0},m^{2}_{f_i},m^{2}_{f_i},m^{2}_{f_i})  \bigg|^{2},
\end{eqnarray}

%\begin{eqnarray}
%\Gamma(A_0\to W^+ W^-) &=&\frac{g^4}{512 \pi^5 m_W \sqrt{\left(\tau_w^2-4\right)}} \bigg[ |\sum_{i,j} m_i \:g_{fifi}\:\: g_{fifj}^2 \left(B_1-B_2+C_0 \left(-m_i^2+m_j^2+m_W^2\right)\right)|^2\nonumber\\
%&+& |\sum_{i',j'} m_j \:g_{fi'fi'}\:\: g_{fi'fj'}^2 \left(B_1-B_2+C_0 \left(-m_{i'}^2+m_{j'}^2+m_W^2\right)\right)|^2\bigg]
%\end{eqnarray}
%
% donde $i=t,T,T_5,T_6; j=b,B$ y $i'=b, j'=t,T_5,T_6$.

\begin{eqnarray} \label{w2W}
\Gamma(A_0\to W W) &=& \frac{g^4}{512 \pi^5 m_{A_0}} \left(1-\frac{4 m^{2}_{W}}{m^{2}_{A_0}} \right)^{-1/2} \nonumber \\
&\times & \bigg|\sum_{i,j}  m_{f_i} \:g_{A_0 f_i f_i} \bigg( \left(-m_{f_i}^2+m_{b_j}^2+m_W^2\right) C_0(m^{2}_{A_0},m^{2}_{W},m^{2}_{W},m^{2}_{f_i},m^{2}_{f_i},m^{2}_{b_j})  \nonumber\\ 
&+&  B_0(m^{2}_{A_0},m^{2}_{f_i},m^{2}_{f_i}) - B_0(m^{2}_{W},m^{2}_{b_j},m^{2}_{f_i})  \bigg) \nonumber \\
&+& \sum_{i,j}  m_{b_j} \:g_{A_0 b_j b_j} \bigg( \left(-m_{b_j}^2+m_{f_i}^2+m_W^2\right) C_0(m^{2}_{A_0},m^{2}_{W},m^{2}_{W},m^{2}_{b_j},m^{2}_{b_j},m^{2}_{f_i})  \nonumber\\ 
&+&  B_0(m^{2}_{A_0},m^{2}_{b_j},m^{2}_{b_j}) - B_0(m^{2}_{W},m^{2}_{f_i},m^{2}_{b_j})  \bigg) \bigg|^{2}.
\end{eqnarray}

\noindent In Eqs.~(\ref{wff})-(\ref{w2W}), the scalar Passarino-Veltman functions are represented by
$C_0(0,0,m_{A_0}^2,m_{f_i}^2,m_{f_i}^2,m_{f_i}^2)$, $C_0(m_{A_0}^2,m^{2}_{Z},0, m_{f_i}^2,m_{f_i}^2,m_{f_i}^2)$, $C_0(m^{2}_{A_0},m^{2}_{Z},m^{2}_{Z},m^{2}_{f_i},m^{2}_{f_i},m^{2}_{f_i})$, $C_0(m^{2}_{A_0},m^{2}_{W},m^{2}_{W},m^{2}_{f_i},m^{2}_{f_i},m^{2}_{b_j})$,   $B_0(m_{A_0}^2, m_{f_i}^2,m_{f_i}^2)$, $ B_0(m_{Z}^2, m_{f_i}^2,m_{f_i}^2)$, $B_0(m^{2}_{W},m^{2}_{b_j},m^{2}_{f_i})$,  $B_0(m^{2}_{A_0},m^{2}_{b_j},m^{2}_{b_j})$, and $B_0(m^{2}_{W},m^{2}_{f_i},m^{2}_{b_j})$.
On the other hand, the effective couplings ($ g_{\gamma f_i f_i}$, $g_{A_0 f_i f_i}$, $g_{A_0 b_j b_j}$, $g^{Z f_i f_i}_{A}$, and $g^{Z f_i f_i}_{V}$) are explicitly provided in Appendix~\ref{rulesF}.

\section{Numerical results} \label{results}

In this section, we present the results for the  decay widths and branching ratios of the  $A_0 \to $ $bb$, $tt$,  $\gamma tt$,
 $\gamma bb$,  $Ztt$, $Zbb$, $Wtb$,  $h_0 tt$, $h_0 bb$,  $WW$,  $gg$, $ZZ$, $\gamma\gamma$,  $\gamma Z$ processes.
We also present our  numerical analysis for the production via gluon fusion of the pseudoscalar $A_0$  at current and future colliders within the BLHM framework.

The BLHM parameters involved in our calculation are $m_{A_0}$, $\tan \beta$, and $f$. In this way, to establish a benchmark scenario, the  pseudoscalar production cross section was calculated by setting the pseudoscalar mass to $m_{A_0}=500$ GeV~\cite{ATLAS:2024jja,ATLAS:2022rws,ATLAS:2020gxx,CMS:2019ogx}. Given the value of $m_{A_0}$, we can determine the allowed range of values for the parameter $\tan \beta$ (the ratio of the
vacuum expectation values of the two Higgs doublets), which is defined by the following equation~\cite{JHEP09-2010,Kalyniak:2013eva}:

 \begin{eqnarray}\label{betalimit}
1 < &\text{tan} & \beta  <  \sqrt{ \frac{2+2 \sqrt{\big(1-\frac{m^{2}_{h_0} }{m^{2}_{A_0}} \big) \big(1-\frac{m^{2}_{h_0} }{4 \pi v^{2}}\big) } }{ \frac{m^{2}_{h_0}}{m^{2}_{A_0}} \big(1+ \frac{m^{2}_{A_0}- m^{2}_{h_0}}{4 \pi v^{2}}  \big) } -1 }.
\end{eqnarray}
 
 \noindent With respect to the energy scale $f$, this parameter characterizes the scale of new physics and plays a central role in the generation of the masses of the new heavy quarks. Constraints on $f$ arise both from theoretical considerations, such as avoiding excessive fine-tuning in the heavy quark sector,  and from experimental bounds on the production of these quarks: $f\in [1000,2000]$ GeV~\cite{JHEP09-2010,Kalyniak:2013eva,Godfrey:2012tf,Cruz-Albaro:2024vjk}.
An additional set of parameters involved in our analysis are the Yukawa couplings $y_i$ ($i=1,2,3$), which are assigned the values $y_1=0.61$, $y_2=0.84$, and $y_3=0.35$~\cite{Martinez-Martinez:2024lez,Cruz-Albaro:2024vjk,Cruz-Albaro:2023pah,Cruz-Albaro:2022kty} .

\begin{figure}[H]
\center
\subfloat[]{\includegraphics[width=8.10cm]{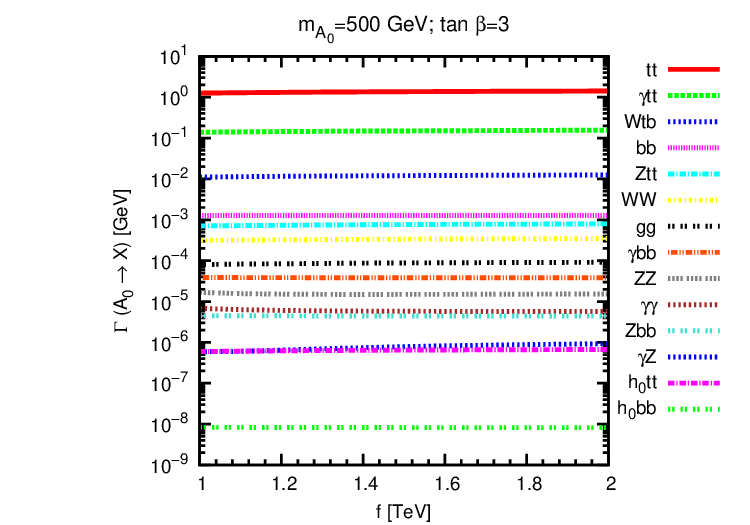}}
\subfloat[]{\includegraphics[width=8.10cm]{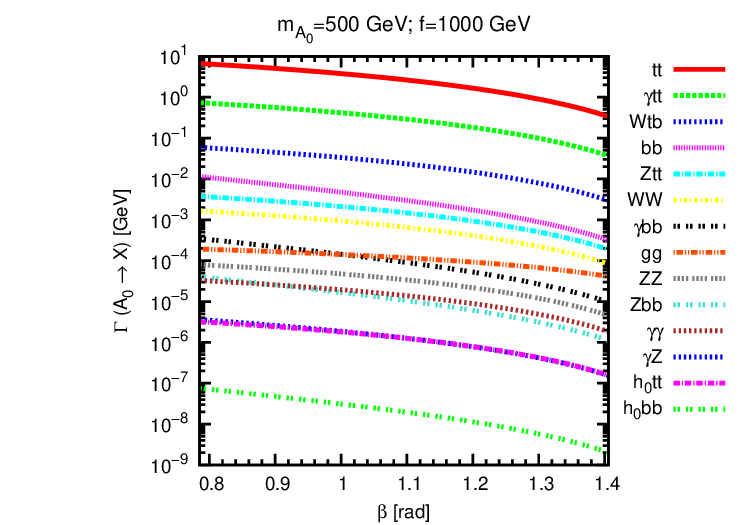}}
\caption{ \label{anchura-fbeta} Decay widths for the $A_0\to X$ processes where $X \equiv $ $tt$, $bb$, $\gamma tt$, $\gamma bb$,  $Ztt$, $Zbb$, $Wtb$,  $h_0 tt$, $h_0 bb$,  $WW$,  $gg$, $ZZ$, $\gamma\gamma$,  $\gamma Z$. a) $\Gamma(A_0\to X)$ as a function of the $f$ energy scale. 
b)  $\Gamma(A_0\to X)$ as a function of the parameter $\beta$ ($\tan \beta \in [1,6]$).}
\end{figure}

We first investigate how variations in the parameters $f$ or $\beta$ affect the decay width of the $A_0 \to X$ ($X \equiv $ $tt$, $bb$, $\gamma tt$, $\gamma bb$,  $Ztt$, $Zbb$, $Wtb$,  $h_0 tt$, $h_0 bb$,  $WW$,  $gg$, $ZZ$, $\gamma\gamma$,  $\gamma Z$) processes. 
In Fig.~\ref{anchura-fbeta}(a), we illustrate the behavior of the decay width $\Gamma(A_0 \to X) $ as a function of the scale $f$, while keeping $\tan \beta$ fixed at 3.0.
From this figure, we observe that the largest contributions to $\Gamma(A_0 \to X) $ arise from the two-body and three-body tree-level pseudoscalar decays $\Gamma(A_0 \to t\bar{t})$ and  $\Gamma(A_0 \to \gamma t\bar{t})$, which represent the dominant and subdominant contributions, respectively, throughout the entire energy interval of the $f$ scale analysis:
 $\Gamma(A_0 \to t\bar{t})=[1.26,1.42] $ GeV and  $\Gamma(A_0 \to \gamma t\bar{t})=[1.39,1.56]\times 10^{-1} $ GeV.
  In contrast, the $A_0 \to h_0 b\bar{b}$ decay provides the smallest contribution, i.e., $\Gamma(A_0 \to  h_0 b\bar{b} )=[8.34,8.21]\times 10^{-9} $ GeV while $f\in [1000, 2000]$ GeV.  It is important to note that $A_0 \to X$ decays arising at one-loop level generate numerical contributions of the order of $10^{-4}-10^{-7}$ GeV.
  On the other hand, Fig.~\ref{anchura-fbeta}(b) shows our analysis of the impact of the $\beta$ parameter on $\Gamma(A_0 \to X) $, with the $f$ energy scale fixed at 1000 GeV. In this scenario, the curves that generate the largest contributions to the decay width $\Gamma(A_0 \to X) $ are again given by the tree-level pseudoscalar decays  $A_0 \to t\bar{t} $ and  $A_0 \to \gamma t\bar{t} $: 
   $\Gamma(A_0 \to t\bar{t})=[6.70,3.39\times 10^{-1}] $ GeV and  $\Gamma(A_0 \to \gamma t\bar{t})=[7.36 \times 10^{-1},3.72\times 10^{-2}] $ GeV when $\beta \in [\tan^{-1} (1), \tan^{-1} (6)]$ rad.
   Additionally, the $A_0 \to h_0 b\bar{b}$ decay provides the most suppressed contribution, $\Gamma(A_0 \to  h_0 b\bar{b} )=[7.57 \times 10^{-8}, 2.03\times 10^{-9}] $ GeV.
   As for the $A_0$ pseudoscalar decays arising at one-loop level, in this context the processes generate contributions of the order of $10^{-3}$ to $10^{-7}$ GeV, these contributions are slightly larger compared to those obtained in the $\Gamma(A_0 \to X) $ vs. $f$ analysis scenario.
   As can be seen clearly in Figs.~\ref{anchura-fbeta}(a) and~\ref{anchura-fbeta}(b), $\Gamma(A_0 \to X) $ depends mainly on the parameter $\beta$. This effect is evident in the curves of $\Gamma(A_0 \to X) $ vs. $\beta$, which show a pronounced decrease of about one order of magnitude for larger values of $\beta$. In contrast, $\Gamma(A_0 \to X) $ shows little sensitivity to changes in $f$ scale values; the corresponding curves decrease slightly as $f$ increases.

\begin{figure}[H]
\center
\subfloat[]{\includegraphics[width=8.10cm]{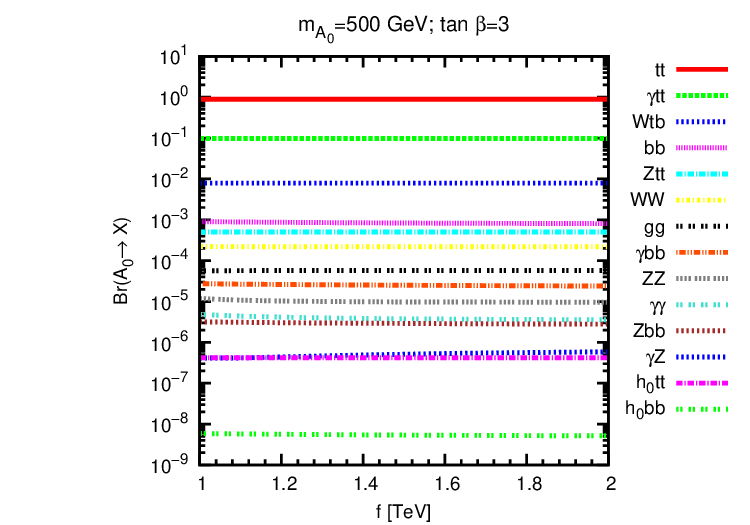}}
\subfloat[]{\includegraphics[width=8.10cm]{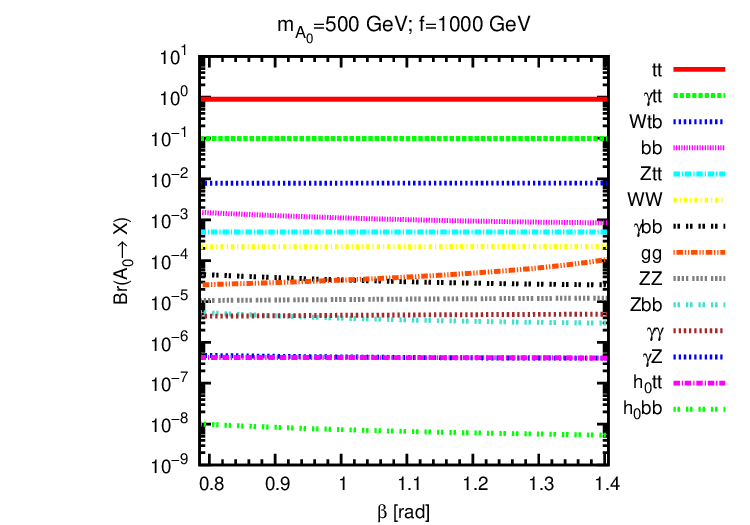}}
\caption{ \label{Brf} Branching ratios for the $A_0\to X$ processes where $X \equiv $ $tt$, $bb$, $\gamma tt$, $\gamma bb$,  $Ztt$, $Zbb$, $Wtb$,  $h_0 tt$, $h_0 bb$,  $WW$,  $gg$, $ZZ$, $\gamma\gamma$,  $\gamma Z$. a) $\text{Br}(A_0\to X)$ as a function of the $f$ energy scale.
b) $\text{Br}(A_0\to X)$ as a function of the  parameter $\beta$ ($\tan \beta \in [1,6]$).}
\end{figure}

Fig.~\ref{Brf} shows the corresponding branching ratios $\text{Br}( A_0 \to X)$ for the $A_0 \to X$ decays. The total decay width ($\Gamma_{A_0}$) of the pseudoscalar Higgs boson $A_0$ has been calculated taking into account the following pseudoscalar decay modes: $tt$, $bb$, $\gamma tt$, $\gamma bb$,  $Ztt$, $Zbb$, $Wtb$,  $h_0 tt$, $h_0 bb$,  $WW$,  $gg$, $ZZ$, $\gamma\gamma$,  $\gamma Z$.  In the left plot of Fig.~\ref{Brf}, the dependence of 
$\text{Br}(A_0\to X)$ vs. $f$ is analyzed. In this plot, we see that the most probable decay channels of the pseudoscalar are given by the 
$A_0 \to t\bar{t}$  and $A_0 \to \gamma t\bar{t}$ processes when $f\in [1000,2000]$ GeV. These decays arise at tree level and generate dominant and subdominant contributions to $\text{Br}( A_0 \to X)$: $\text{Br}( A_0 \to t\bar{t})\approx 8.92 \times 10^{-1}$ and $\text{Br}( A_0 \to \gamma t\bar{t})\approx 9.81 \times 10^{-2}$, respectively. While the numerical values do not show a strong sensitivity of  $\text{Br}( A_0 \to X)$ to $f$ at first glance, a variation becomes noticeable after a few digits beyond the decimal point. On the other hand, it is observed that the $A_0 \to h_0 b\bar{b}$ decay generates the smallest contribution to the branching ratio, that is,  $\text{Br}( A_0 \to h_0 b\bar{b})= [5.90,5.17]\times 10^{-9}$.
Regarding the right plot of Fig.~\ref{Brf}, in this scenario we generate curves by varying the $\beta$ parameter while keeping the other parameter, $f$, fixed at 1000 GeV.
 In the corresponding figure, we see that the two main contributions are  $\text{Br}( A_0 \to t\bar{t})\approx 8.92 \times 10^{-1}$ and 
 $\text{Br}( A_0 \to \gamma t\bar{t})= [9.80,9.81] \times 10^{-2}$ when $\beta \in [\tan^{-1} (1), \tan^{-1} (6)]$ rad.
On the contrary, the most suppressed contribution is given by the $A_0 \to h_0 b\bar{b}$ decay which gives  branching ratios of 
$\text{Br}( A_0 \to h_0 b\bar{b})=[1.01\times 10^{-8}, 5.35\times 10^{-9}]$ in the study interval for the $\beta$ parameter.
Notably, in both study scenarios ($\text{Br}( A_0 \to X)$ vs. $f$ and $\text{Br}( A_0 \to X)$ vs. $\beta$), the one‑loop  $A_0 \to X$ decays produce branching ratios on the order of $10^{-4}$ to $10^{-7}$.

\subsection{Production cross section of the pseudoscalar $A_0$ at the LHC and FCC-hh }

In this subsection, we present the Breit-Wigner resonance formula, which is useful for describing the resonant production of the pseudoscalar Higgs boson $A_0$. Although Eq.~(\ref{BW}) is determined just at the resonance of the pseudoscalar, it provides an approximate description of the pseudoscalar production mechanism via gluon fusion in the BLHM, it can also serve as a theoretical reference for guiding experimental searches for this new particle. Under this approximation, the production cross section of $A_0$ via gluon fusion can be expressed as follows~\cite{pdg:2023}:

\begin{eqnarray} \label{BW}
  \sigma(gg\to A_0\to Y) &=& \frac{\pi}{36}\frac{\Gamma(A_0\to gg) \Gamma(A_0\to Y)}{m_{A_0}^2 \Gamma^{2}_{A_0}},
\end{eqnarray}

\noindent   where $\Gamma(A_0\to gg)$  and $\Gamma(A_0\to Y)$ (with $Y \equiv \gamma\gamma, \gamma Z, ZZ, gg, WW$) are the partial decay widths for an on-shell $A_0$ to decay to the initial and final states, respectively.

\begin{figure}[H]
\center
\subfloat[]{\includegraphics[width=8.10cm]{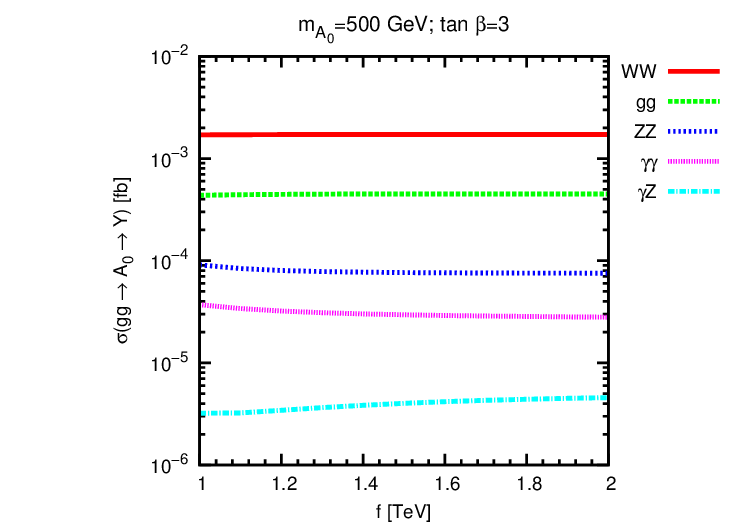}}
\subfloat[]{\includegraphics[width=8.10cm]{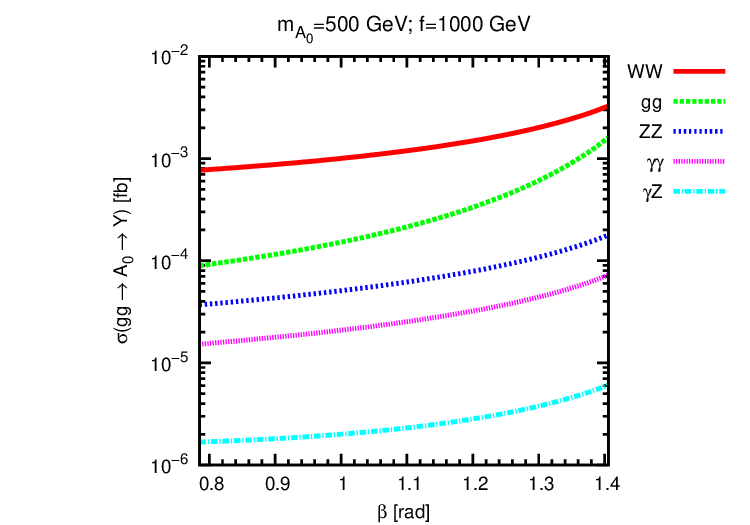} }
\caption{ \label{sigmafbeta}  Production cross section of the  pseudoscalar $A_0$ (for $m_{A_0}=500$ GeV) via gluon fusion. a) $\sigma(gg \to A_0 \to Y)$ as a function of the $f$ energy scale.
b)  $\sigma(gg \to A_0 \to Y)$ as a function of the  parameter $\beta$ ($\tan \beta \in [1,6]$).}
\end{figure}

Using Eq.~(\ref{BW}), we generated the curves shown in Fig.~\ref{sigmafbeta} for $m_{A_0}=500$ GeV, in which we studied the behavior of the  cross section
 $\sigma(gg \to A_0 \to Y)$ as a function of parameter $f$ or $\beta$, while fixing the other model parameters. We first discuss the behavior observed in Fig.~\ref{sigmafbeta}(a) when $\tan \beta=3.0$. 
  Note that the magnitude of $\sigma(gg \to A_0 \to Y)$ depends slightly on the $f$ energy scale, except for the $gg \to A_0 \to \gamma Z$ decay channel, which gives a curve with more appreciable changes within the range of study of the $f$ parameter, $f \in[1000,2000]$ GeV.
 In the same context, we find that the  $gg \to A_0 \to W W$ and  $gg \to A_0 \to \gamma Z$ decay channels generate the largest and smallest cross sections over the entire $f$ interval: $\sigma(gg \to A_0 \to W W)=[1.70,1.71]\times 10^{-3}$ fb and $\sigma(gg \to A_0 \to \gamma Z)=[3.21,4.57]\times 10^{-6}$ fb.
 Regarding Fig.~\ref{sigmafbeta}(b), this plot shows the results for the pseudoscalar production cross section when the  $f$ scale is fixed at $f=1000$ GeV, and the $\beta$ parameter  varies within the range $\tan^{-1}(1)$ to $\tan^{-1}(6)$ rad. Once again, the  $gg \to A_0 \to W W$ process gives the largest contribution to $\sigma(gg \to A_0 \to Y)$, while the $gg \to A_0 \to \gamma Z$ decay channel yields the most suppressed contribution, i.e.,
  $\sigma(gg \to A_0 \to W W)=[7.68\times 10^{-4},3.26 \times 10^{-3}]$ fb and $\sigma(gg \to A_0 \to \gamma Z)=[1.68,6.06]\times 10^{-6}$ fb.  
 The various curves obtained in the analysis of the $\sigma(gg \to A_0 \to Y)$ production cross section as a function of $\beta$ exhibit a growth of up to two orders of magnitude as  $\beta$ increases toward $1.4$ rad, as shown in the Fig.~\ref{sigmafbeta}(b). It is evident that the region with the greater predictive importance  corresponds to values of  $\beta$ around 1.4 rad.

\begin{figure}[H]
\center
\includegraphics[width=8.10cm]{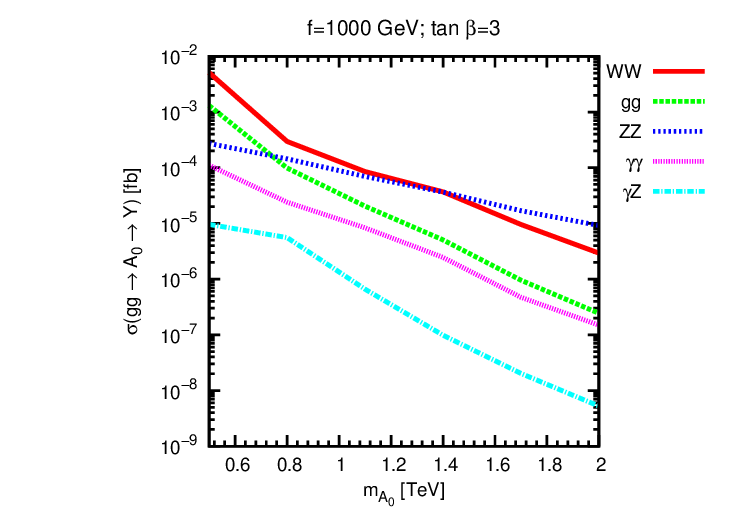}
\includegraphics[width=8.10cm]{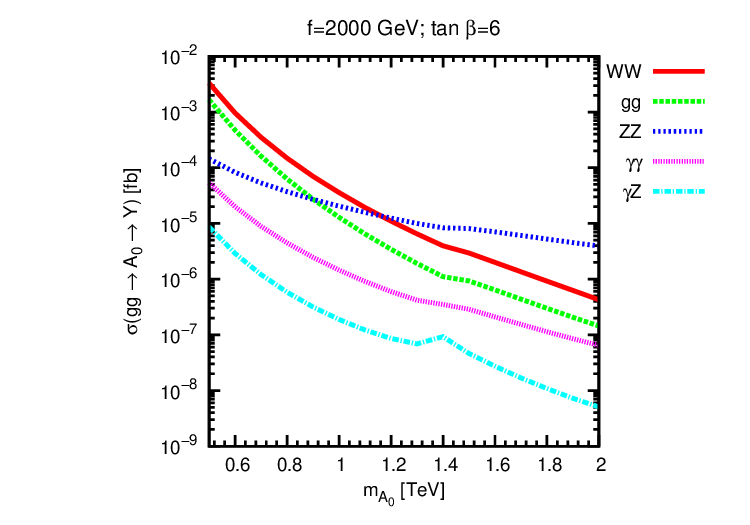}
\caption{ \label{sigmamA0}
 Production cross section of the pseudoscalar $A_0$  via gluon fusion as a function of the  parameter $m_{A_0}$: a) $f=1000$ GeV and $\tan \beta=3$, and  
 b)  $f=2000$ GeV and $\tan \beta=6$.}
\end{figure}

We now turn to discuss the dependence of $\sigma(gg \to A_0 \to Y)$ on the $m_{A_0}$ parameter when it varies from 500 to 2000 GeV, as depicted in Fig.~\ref{sigmamA0}.
 In this figure, the curves have been generated for the values of $f = 1000$ GeV and $\tan \beta = 3$, as well as $f = 2000$ GeV and $\tan \beta = 6$.
  In the first scenario (see Fig.~\ref{sigmamA0}(a)), numerical evaluation shows that the $gg \to A_0 \to WW$  decay channel provides the largest cross section when $m_{A_0}$ is  approximately between 500 and 1360 GeV, $\sigma(gg \to A_0 \to W W)=[5.10\times 10^{-3},3.69 \times 10^{-5}]$ fb. Outside this range, the largest contribution comes from the $gg \to A_0 \to ZZ$ process with $\sigma(gg \to A_0 \to ZZ)=[3.69 \times 10^{-5},9.17\times 10^{-6}]$ fb when  $m_{A_0}\in [1360,2000]$ GeV.
  With respect to the second scenario (see Fig.~\ref{sigmamA0}(b)), again the $gg \to A_0 \to WW$ and $gg \to A_0 \to ZZ$ decay channels generate the largest contributions to $\sigma(gg \to A_0 \to Y)$, i.e., $\sigma(gg \to A_0 \to WW)=[3.30\times 10^{-3}, 1.94 \times 10^{-5}]$ fb and
   $\sigma(gg \to A_0 \to ZZ)=[1.94 \times 10^{-5},3.93\times 10^{-6}]$ fb when $m_{A_0} \in[500,1160]$ GeV and $m_{A_0}\in [1160,2000]$ GeV, respectively. 
 In both study scenarios, we observe that the pseudoscalar production cross section decreases as the mass of  $A_0$ increases.  It is also evident that $\sigma(gg \to A_0 \to Y)$  strongly depends on the  $m_{A_0}$ parameter. In general, the $gg \to A_0 \to \gamma Z$ decay channel provides the smallest numerical contributions to  $\sigma(gg \to A_0 \to Y)$.

In the experimental context, the production mechanism of the pseudoscalar  $A_0$ via gluon fusion could be studied at the LHC, particularly in its future upgrades such as the High-Luminosity LHC (HL-LHC) and the High-Energy LHC (HE-LHC), as well as in future high-energy colliders like the FCC-hh. The HL-LHC~\cite{ZurbanoFernandez:2020cco,Azzi:2019yne,Cepeda:2019klc,FCC:2018bvk} is planned to operate at a center-of-mass energy of 14 TeV with an integrated luminosity of 3000 fb$^{-1}$, while the HE-LHC~\cite{ZurbanoFernandez:2020cco,Azzi:2019yne,Cepeda:2019klc,FCC:2018bvk} would provide proton-proton collisions at a center-of-mass energy of 27 TeV with an integrated luminosity of 10000 fb$^{-1}$. Accurate exploration of rare processes will be made possible by the LHC upgrades, which promise to increase the collision energy.
As for the future hadron collider, the FCC-hh is designed to operate at a center-of-mass energy of $\sqrt{s}=100$ TeV and to collect a total integrated luminosity of 30000 fb$^{-1}$~\cite{MammenAbraham:2024gun,Mangano:2022ukr,FCC:2018byv,FCC:2018vvp}. This would allow the exploration of regions of parameter space that are currently inaccessible and would significantly increase the potential to study processes with extremely small cross sections. The FCC-hh is considered one of the most promising projects for the discovery of new particles in the coming decades.

Using the numerical results for the pseudoscalar production cross sections and considering the expected integrated luminosities of the HL-LHC, HE-LHC, and FCC-hh colliders, we can estimate the number of events that could be observed at these colliders for the  $A_0 \to Y$  pseudoscalar decays. Tables~\ref{WW}-\ref{gammaZ} present the expected number of events associated with the  $A_0 \to Y$ decays when the new physics scale $f$ takes specific values such as 1000 and 2000 GeV. It is important to emphasize that these event estimates have been generated  by setting $m_{A_0}=500$ GeV.
 In this context, we find that the production of the pseudoscalar $A_0$ through the  $A_0 \to WW$ and $A_0 \to gg$ decays (see Tables~\ref{WW} and~\ref{gg}), which arise at one-loop level, provides very promising scenarios for the experiments of interest.

\begin{table}[H]
\begin{center}
\caption{The number of expected events related to $A_0\to WW$ decay.
\label{WW}}
\begin{tabular}{|c|c|c|c|c|c|}\hline\hline
\multirow{3}{*}{ } & \multicolumn{3}{c|}{$m_{A_0}=500$ GeV, $\tan \beta=6$} \\
\cline{2-4}
\multirow{3}{*}{$f$ [TeV]} & \multicolumn{3}{c|}{\textbf{No. of expected events at the colliders:}} \\
  \cline{2-4}
 & HL-LHC   & HE-LHC & FCC-hh \\
 & $\mathcal{L}=3\, 000$ fb$^{-1}$   & $\mathcal{L}=10\, 000$ fb$^{-1}$ & $\mathcal{L}=30\, 000  $ fb$^{-1}$ \\
\hline
1 & 10 &  32  &  98   \\
2 & 10  & 33  &  99  \\
\hline
\hline
\end{tabular}
\end{center}
\end{table}

\begin{table}[H]
\begin{center}
\caption{The number of expected events related to $A_0\to gg$ decay.
\label{gg}}
\begin{tabular}{|c|c|c|c|c|c|}\hline\hline
\multirow{3}{*}{ } & \multicolumn{3}{c|}{$m_{A_0}=500$ GeV, $\tan \beta=6$} \\
\cline{2-4}
\multirow{3}{*}{$f$ [TeV]} & \multicolumn{3}{c|}{\textbf{No. of expected events at the colliders:}} \\
  \cline{2-4}
 & HL-LHC   & HE-LHC & FCC-hh \\
 & $\mathcal{L}=3\, 000$ fb$^{-1}$   & $\mathcal{L}=10\, 000$ fb$^{-1}$ & $\mathcal{L}=30\, 000  $ fb$^{-1}$ \\
\hline
1 & 5 &  16  &  48   \\
2 & 5  & 17  &  50  \\
\hline
\hline
\end{tabular}
\end{center}
\end{table}

\begin{table}[H]
\begin{center}
\caption{The number of expected events related to $A_0\to ZZ$ decay.
\label{ZZ}}
\begin{tabular}{|c|c|c|c|c|c|}\hline\hline
\multirow{3}{*}{ } & \multicolumn{3}{c|}{$m_{A_0}=500$ GeV, $\tan \beta=6$} \\
\cline{2-4}
\multirow{3}{*}{$f$ [TeV]} & \multicolumn{3}{c|}{\textbf{No. of expected events at the colliders:}} \\
  \cline{2-4}
 & HL-LHC   & HE-LHC & FCC-hh \\
 & $\mathcal{L}=3\, 000$ fb$^{-1}$   & $\mathcal{L}=10\, 000$ fb$^{-1}$ & $\mathcal{L}=30\, 000  $ fb$^{-1}$ \\
\hline
1 & 1 &  2  &  5   \\
2 & 0  & 1  &  4  \\
\hline
\hline
\end{tabular}
\end{center}
\end{table}

\begin{table}[H]
\begin{center}
\caption{The number of expected events related to $A_0\to \gamma \gamma$ decay.
\label{2gamma}}
\begin{tabular}{|c|c|c|c|c|c|}\hline\hline
\multirow{3}{*}{ } & \multicolumn{3}{c|}{$m_{A_0}=500$ GeV, $\tan \beta=6$} \\
\cline{2-4}
\multirow{3}{*}{$f$ [TeV]} & \multicolumn{3}{c|}{\textbf{No. of expected events at the colliders:}} \\
  \cline{2-4}
 & HL-LHC   & HE-LHC & FCC-hh \\
 & $\mathcal{L}=3\, 000$ fb$^{-1}$   & $\mathcal{L}=10\, 000$ fb$^{-1}$ & $\mathcal{L}=30\, 000  $ fb$^{-1}$ \\
\hline
1 & 0 &  1  &  2  \\
2 & 0  & 0  &  2  \\
\hline
\hline
\end{tabular}
\end{center}
\end{table}

\begin{table}[H]
\begin{center}
\caption{The number of expected events related to $A_0\to \gamma Z$ decay.
\label{gammaZ}}
\begin{tabular}{|c|c|c|c|c|c|}\hline\hline
\multirow{3}{*}{ } & \multicolumn{3}{c|}{$m_{A_0}=500$ GeV, $\tan \beta=6$} \\
\cline{2-4}
\multirow{3}{*}{$f$ [TeV]} & \multicolumn{3}{c|}{\textbf{No. of expected events at the colliders:}} \\
  \cline{2-4}
 & HL-LHC   & HE-LHC & FCC-hh \\
 & $\mathcal{L}=3\, 000$ fb$^{-1}$   & $\mathcal{L}=10\, 000$ fb$^{-1}$ & $\mathcal{L}=30\, 000  $ fb$^{-1}$ \\
\hline
1 & 0 &  0  &  0  \\
2 & 0  & 0  &  0  \\
\hline
\hline
\end{tabular}
\end{center}
\end{table}

Additionally, Table~\ref{WW1000} presents the number of events for the production of the pseudoscalar $A_0$ through the $A_0 \to Y$  decays when 
$m_{A_0}=1000$ GeV is selected. It is important to mention that, for this benchmark scenario, only the FCC-hh appears to be within reach for the potential discovery of this new particle.  The production of the pseudoscalar $A_0$ via gluon fusion would only be possible through the $A_0 \to WW$, $A_0 \to gg$, and $A_0 \to ZZ$ decays.

\begin{table}[H]
\begin{center}
\caption{The number of expected events at the FCC-hh for the $A_0 \to Y$ decay.
\label{WW1000}}
\begin{tabular}{|c|c|c|c|c|c|c|}\hline\hline
\multirow{3}{*}{ } & \multicolumn{5}{c|}{$m_{A_0}=1000$ GeV, $\tan \beta=6$} \\
\cline{2-6}
\multirow{3}{*}{$f$ [TeV]} & \multicolumn{5}{c |}{\textbf{No. of expected events at the FCC-hh:}} \\
  \cline{2-6}
 & $A_0\to WW$   & $A_0\to gg$ & $A_0\to ZZ$  & $A_0\to \gamma \gamma$ & $A_0\to \gamma Z$  \\
 & $\mathcal{L}=30 000$ fb$^{-1}$   & $\mathcal{L}=30 000$ fb$^{-1}$ & $\mathcal{L}=30 000  $ fb$^{-1}$  & $\mathcal{L}=30 000  $ fb$^{-1}$  & $\mathcal{L}=30 000  $ fb$^{-1}$ \\
\hline
1 & 1 &  1  &  1 & 0  & 0  \\
2 & 1  & 0  &  0  & 0  & 0 \\
\hline
\hline
\end{tabular}
\end{center}
\end{table}

\section{Conclusions} \label{conclusions}

Within the BLHM framework, we investigate the production of the pseudoscalar $A_0$ through gluon fusion at future colliders, such as the HL-LHC, HE-LHC, and FCC-hh. These experiments promise to extend current searches for heavy particles, such as the pseudoscalar of interest to us.

In our study of the pseudoscalar $A_0$, we analyze the impact of the free parameters of the BLHM ($f$, $\beta$, and $m_{A_0}$) on the partial decay widths and branching fractions of the pseudoscalar, considering both two-and three-body decays at tree level, as well as one-loop level decays. For the one-loop decays, we account for the virtual particle effects induced by both the BLHM and SM particles.
We also investigate the resonant production of the pseudoscalar $A_0$ using the Breit–Wigner resonance formula, evaluated specifically at the resonance peak of the pseudoscalar. Within this approach, we examine the sensitivity of the $\sigma(gg \to A_0 \to Y)$  production cross section to the parameters $f$, $\beta$, and $m_{A_0}$, considering the following study intervals: $f \in [1000,2000]$ GeV,  $\beta \in [\tan^{-1}(1),\tan^{-1}(6)]$ radians, and  $m_{A_0} \in [500,2000]$ GeV.
Our numerical results show that $\sigma(gg \to A_0 \to Y)$  increases by up to two orders of magnitude as $\beta$ approaches 1.4 radians, indicating a strong dependence on this parameter. We find that the region of highest predictive significance corresponds to values of $\beta$ around approximately 1.4 radians. Additionally, we observe that the pseudoscalar production cross section decreases as the mass $m_{A_0}$ increases.

To establish a benchmark, Tables~\ref{WW}-\ref{gammaZ}  present the expected number of events from the production of the pseudoscalar for two different mass values: $m_{A_0}=500$ GeV and $m_{A_0}=1000$ GeV. To estimate the number of events that could be observed at the colliders for the pseudoscalar decays
 $A_0 \to Y$, we have considered the expected integrated luminosities of the HL-LHC, HE-LHC, and FCC-hh.
In the first scenario, when $m_{A_0}=500$ GeV, we find that the production of the pseudoscalar $A_0$ through the $A_0 \to WW$ and $A_0 \to gg$  decay channels (see Tables~\ref{WW} and ~\ref{gg}), which arise at one-loop level, provide very promising scenarios for the experiments of interest.
In the second scenario, when  $m_{A_0}=1000$ GeV, only the FCC-hh appears to be within reach for the potential discovery of the pseudoscalar Higgs boson $A_0$.  Furthermore, the study of the pseudoscalar $A_0$ at resonance is an excellent starting point for the search for new physics. Our study complements other studies in the context of present and future hadron-hadron, hadron-lepton, and lepton-lepton colliders on the production and decay of the pseudoscalar $A_0$.

\vspace{6.0cm}

\begin{center}
{\bf Acknowledgements}
\end{center}

 E. Cruz-Albaro appreciates the postdoctoral stay at the
Universidad Autónoma de Zacatecas, M\'exico. A.G.R. and M.A.H.R. thank SNII and PROFEXCE (M\'exico).

\vspace{1cm}
%\newpage

\begin{center}
   {\bf Declarations}
\end{center}

 Data Availability Statement: All data generated or analyzed during this study are included in this article.

%\vspace{2cm}

\newpage

\appendix

\section{Effective couplings in the BLHM} \label{rulesF}

In this Appendix, we present the effective couplings involved in our calculation of the production of the pseudoscalar $A_0$.

\begin{table}[H]
\caption{Effective couplings involved in our calculation of the decay of the pseudoscalar boson $A_0$.
\label{FeyRul0}}
%\begin{tabular}{|c | p{14.3cm}|}
\begin{tabular}{|p{2.5cm}| p{12.5cm}|}
\hline
\hline
      &   \hspace{3.5cm}   \textbf{Effective couplings} \\
\hline
\hline

$g_{A_0 \, t \bar{t}} $ &  $ \frac{3 c_\beta y_1 y_2 y_3 }{\sqrt{y_1^2 + y_2^2} \sqrt{y_1^2 + y_3^2}}-\frac{2 c_\beta s_\beta v y_2 y_3  (y_3^2 - 2 y_1^2)}{f \sqrt{y_1^2 + y_2^2} (y_1^2 + y_3^2)}-\frac{3 c_\beta s_\beta^2 v^2 y_1 y_2 y_3^3 }{2 f^2 (y_1^2 + y_2^2)^{5/2} (y_1^2 + y_3^2)^{5/2}} \bigg( 4 y_1^6 + 4(2 y_2^2 + y_3^2)y_1^4 + y_1^2 (4y_2^4 + 8 y_3^2 y_2^2 - 2\sqrt{y_1^2 + y_3^2})  + y_2^2 (4 y_2^2 y_3^2 + \sqrt{y_1^2 + y_3^2}) \bigg)$\\

\hline
\hline
 $g_{A_0 \, b \bar{b}} $ & $-c_\beta y_b $ \\
 \hline
 \hline
 $g_{A_0 \, T \bar{T}} $ & $-\frac{c_\beta s_\beta v y_1^2 }{f (y_1^2 + y_2^2)^{3/2} (y_1^2 + y_3^2) (y_2^2 - y_3^2)} \bigg( y_1^4 (y_2^2 + 2 y_3^2) + y_1^2 (8 y_2^2 y_3^2 - 2 y_3^4) - y_2^2 (y_2^4 + 3 y_2^2 y_3^2 - 7 y_3^4)\bigg)$ \\
\hline
\hline
 $g_{A_0 \, T_{5} \bar{T_{5}}} $ & $ -\frac{9 c_\beta s_\beta v y_1^2 y_2^2 y_3^2}{f \left(y_1^2 + y_3^2 \right)^{\frac{3}{2}} \left( y_3^2 - y_2^2 \right)}$ \\
\hline
\hline
 $g_{A_0 \, T_{6} \bar{T}_{6}} $ & $\frac{c_\beta s_\beta v y_1}{f}$ \\
\hline
\hline
 $g_{A_0 \, T^{23} \bar{T}^{23}} $ & $\frac{c_\beta s_\beta v y_1 \left(3 y_1^2 - 5 y_3^2\right)}{f\left( y_1^2 + y_3^2\right)} $ \\
\hline
\hline
$g_{A_0 A_0 h_0} $ & $-\frac{1}{2}v\lambda_0 sin(\alpha-\beta) sin(2\beta) $ \\
\hline
\hline
\end{tabular}
\end{table}

\begin{table}[H]
\caption{Vector and axial-vector coupling constants of the interaction vertex involving a gauge boson with a pair of fermions.
\label{FeyRul0}}
%\begin{tabular}{|c | p{14.3cm}|}
\begin{tabular}{|p{2.5cm}| p{4.5cm}| p{9.5cm}|c|}
\hline
\hline
  \textbf{Vertex}    &   \hspace{0.5cm}\textbf{Vector couplings} & \hspace{0.5cm} \textbf{Axial-vector couplings}\\
\hline
\hline

$\gamma t\bar{t}$ & $g_{\gamma t\bar{t}}^{V}= \frac{2 g s_W }{3} $ & $g_{\gamma t\bar{t}}^{A}=0$\\
        \hline
$\gamma b\bar{b}$      & $g_{\gamma b\bar{b}}^{V}= -\frac{ g s_W}{3}$  &$g_{\gamma b\bar{b}}^{A}=0$\\
        \hline
$\gamma T\bar{T}$      & $g_{\gamma T\bar{T}}^{V}= \frac{2g s_W}{3}$ &  $g_{\gamma T\bar{T}}^{A}=\frac{g s_W v^2}{3f^2}\bigg(c_\beta^2-   \frac{y_1^2(4c_\beta^2(y_2^2-y_3^2)^2+s_\beta^2(2y_2^2+y_3^2)^2)}{(y_1^2+y_2^2)(y_2^2-y_3^2)^2} \hfill \break  +\frac{s_\beta^2y_3^2(y_2^2-2y_1^2)^2}{(y_1^2+y_2^2)^2(y_1^2+y_3^2)} \bigg)$ \\
        \hline
$\gamma T_5\bar{T_5}$     &   $g_{\gamma T_5\bar{T_5}}^{V}= \frac{2}{3} g s_W $ & $g_{\gamma T_5\bar{T_5}}^{A}=\frac{3 gs_\beta^2s_W v^2 y_1^2 y_2^2 \left(y_3^2-y_1^2\right)}{2 f^2\left(y_1^2+y_3^2\right)^2 \left(y_2^2-y_3^2\right)}$ \\
        \hline
$\gamma T_6\bar{T_6}$     &  $g_{\gamma T_6\bar{T_6}}^{V}= \frac{2}{3} g s_W $ & $g_{\gamma T_6\bar{T_6}}^{A}=-\frac{3c_\beta^2 gs_W  v^2 }{4f^2}$\\

        \hline
$\gamma T^{2/3}\bar{T}^{2/3}$     &  $g_{\gamma T^{2/3}\bar{T}^{2/3}}^{V}=\frac{2}{3} g s_W $ &$g_{\gamma T^{2/3}\bar{T}^{2/3}}^{A}=\frac{g s_\beta^2 s_W v^2}{f^2}$\\
        \hline
%$g f_i f_i $     &  $g_{g f_i f_i }^{V}= g_s \frac{1}{2} \lambda_{ij}$ & $g_{g f_i f_i }^{A}=0$ \\
        \hline
\end{tabular}
\end{table}

\begin{table}[H]
\caption{Continuation of Table~\ref{FeyRul0}.
\label{FeyRul1}}
%\begin{tabular}{|c | p{14.3cm}|}
\begin{tabular}{|p{2.5cm}| p{6.0cm}| p{8.0cm}|c|}
\hline
\hline
  \textbf{Vertex}    &   \hspace{0.5cm}\textbf{Vector couplings} & \hspace{0.5cm} \textbf{Axial-vector couplings}\\
\hline
\hline

$Z t\bar{t}$ & $g_{Z t\bar{t}}^{V}=\frac{g}{c_W}(\frac{1}{4}-\frac{2 s_W^2}{3}) $ & $g_{Z t\bar{t}}^{A}=\frac{g}{4c_W} $ \\
        \hline
$Z b\bar{b}$ &  $g_{Z b\bar{b}}^{V}=\frac{ g }{c_W}(-\frac{1}{4}+\frac{s_W^2}{3})$ & $g_{Z b\bar{b}}^{A}=-\frac{g}{4c_W}$\\
        \hline
$Z T\bar{T}$ &  $g_{Z T\bar{T}}^{V}=\frac{g}{6c_W} \bigg(( 3c_W^2-s_W^2)\hfil \break -\frac{4s_W^2v^2y_1^2(4c_\beta^2(y_2^2-y_3^2)^2 + s_\beta^2(2y_2^2+y_2)^2)}{f^2(y_1^2+y_2^2)(y_2^2-y_3^2)^2}\bigg)$ & $g_{Z T\bar{T}}^{A}= 0$\\
        \hline
$Z T_5\bar{T_5}$ & $g_{Z T_5\bar{T_5}}^{V}=-\frac{2 g s_W^2}{3c_W}$ &  $g_{Z T_5\bar{T_5}}^{A}=\frac{3gs_\beta^2 v^2y_1^2y_2^2}{4c_Wf^2(y_1^2+y_3^2)^2(y_2^2-y_3^2)^2} \bigg((s_W^2y_3^4-c_W^2(y_1^2(y_2^2+2y_3^2)-2y_2^2y_3^2+5y_3^4))-c_W^2(y_1^2+y_3^2)(y_2^2-4y_3^2)\bigg) $\\
        \hline
$Z T_6\bar{T_6}$ & $g_{Z T_6\bar{T_6}}^{V}= -\frac{g \left( 8f^2s_W^2-15c_\beta^2 c_W^2v^2 \right)}{12c_Wf^2}$ & $g_{Z T_6\bar{T_6}}^{A}=-\frac{3gc_Wc_\beta^2 g v^2}{4f^2}$ \\
        \hline
$Z T^{2/3}\bar{T}^{2/3}$ & $g_{Z T_{2/3}\bar{T}_{2/3}}^{V}=-\bigg(\frac{5gs_\beta^2s_W^2v^2}{3c_W f^2}+\frac{c_W g}{2}+\frac{7gs_W^2}{6c_W}\bigg)$ &   $g_{Z T_{2/3}\bar{T}_{2/3}}^{A}=-\frac{gs_\beta^2 s_W^2 v^2}{c_W f^2} $ \\
        \hline
$W^+\bar{t}b$ &  $g_{W\bar{t} b}^{V}=\frac{g}{2\sqrt{2}}$ & $g_{W\bar{t}b}^{A}=-\frac{g}{2\sqrt{2}}$ \\
        \hline
$W^+\bar{T}_5b$ &  $g_{W\bar{T}_5 b}^{V}=-\frac{gs_\beta v y_2(2y_1^2-y_3^2)}{2\sqrt{2}f\sqrt{y_1^2+y_2^2}(y_1^2+y_3^2)}$ & $g_{W\bar{T}_5 b}^{A}=\frac{gs_\beta v y_2(2y_1^2-y_3^2)}{2\sqrt{2}f\sqrt{y_1^2+y_2^2}(y_1^2+y_3^2)}$ \\
        \hline
$W^+\bar{T}_6b$ &  $g_{W\bar{T}_6 b}^{V}=-\frac{\sqrt{2} c_\beta v y_2}{2f\sqrt{y_1^2+y_2^2}}$ & $g_{W\bar{T}_5 b}^{A}=\frac{\sqrt{2} c_\beta v y_2}{2f\sqrt{y_1^2+y_2^2}}$ \\
        \hline
$W^+\bar{T}B$ &  $g_{W\bar{T} B}^{V}=\frac{g}{2\sqrt{2}}$ & $g_{W\bar{T}B}^{A}=-\frac{g}{2\sqrt{2}}$ \\
        \hline
$W^+\bar{T}_5B$ &  $g_{W\bar{T}_5 B}^{V}=-\frac{gs_\beta v y_1(2y_2^2+y_3^2)}{2\sqrt{2}f\sqrt{y_1^2+y_2^2}(y_2^2-y_3^2)}$ & $g_{W\bar{T}_5 B}^{A}=\frac{gs_\beta v y_1(2y_2^2+y_3^2)}{2\sqrt{2}f\sqrt{y_1^2+y_2^2}(y_2^2-y_3^2)}$ \\
        \hline
$W^+\bar{T}_6 B$ &  $g_{W\bar{T}_6 B}^{V}=-\frac{\sqrt{2} c_\beta v y_1}{2f\sqrt{y_1^2+y_2^2}}$ & $g_{W\bar{T}_5 B}^{A}=\frac{\sqrt{2} c_\beta v y_1}{2f\sqrt{y_1^2+y_2^2}}$ \\
        \hline
        \hline
\end{tabular}
\end{table}

\newpage


\newpage




\begin{thebibliography}{99}
%
%\cite{ATLAS:2012yve}
\bibitem{ATLAS:2012yve}
G.~Aad \textit{et al.} (ATLAS Collaboration),
%``Observation of a new particle in the search for the Standard Model Higgs boson with the ATLAS detector at the LHC,''
{\it Phys. Lett.} B \textbf{716}, 1 (2012).
%doi:10.1016/j.physletb.2012.08.020

%\cite{CMS:2012qbp}
\bibitem{CMS:2012qbp}
S.~Chatrchyan \textit{et al.} (CMS Collaboration),
%``Observation of a New Boson at a Mass of 125 GeV with the CMS Experiment at the LHC,''
{\it Phys. Lett.} B \textbf{716}, 30 (2012).
%doi:10.1016/j.physletb.2012.08.021



%\cite{CMS:2023ftu}
\bibitem{CMS:2023ftu}
A.~Hayrapetyan \textit{et al.} (CMS Collaboration),
%``Observation of four top quark production in proton-proton collisions at s=13TeV,''
{\it Phys. Lett.}  B \textbf{847}, 138290 (2023).
%doi:10.1016/j.physletb.2023.138290

%\cite{ATLAS:2023ajo}
\bibitem{ATLAS:2023ajo}
G.~Aad \textit{et al.} (ATLAS Collaboration),
%``Observation of four-top-quark production in the multilepton final state with the ATLAS detector,''
{\it Eur. Phys. J.} C \textbf{83}, 496 (2023);
Erratum: {\it Eur. Phys. J.} C \textbf{84}, 156 (2024).
%doi:10.1140/epjc/s10052-023-11573-0

%\cite{Englert:2014uua}
\bibitem{Englert:2014uua}
C.~Englert, A.~Freitas, M.~M.~M\"uhlleitner, T.~Plehn, M.~Rauch, M.~Spira and K.~Walz,
%``Precision Measurements of Higgs Couplings: Implications for New Physics Scales,''
{\it J. Phys.}  G \textbf{41}, 113001 (2014).
%doi:10.1088/0954-3899/41/11/113001

%\cite{Schmaltz:2002wx}
\bibitem{Schmaltz:2002wx}
M.~Schmaltz,
%``Physics beyond the standard model (theory): Introducing the little Higgs,''
{\it Nucl. Phys. B Proc. Suppl.} \textbf{117}, 40 (2003).
%doi:10.1016/S0920-5632(03)01409-9

%
%\cite{Arkani-Hamed:2002ikv}
\bibitem{Arkani-Hamed:2002ikv}
N.~Arkani-Hamed, A.~G.~Cohen, E.~Katz and A.~E.~Nelson,
%``The Littlest Higgs,''
{\it JHEP} \textbf{07}, 034 (2002).
%doi:10.1088/1126-6708/2002/07/034


%\cite{Chang:2003zn}
\bibitem{Chang:2003zn}
S.~Chang,
%``A 'Littlest Higgs' model with custodial SU(2) symmetry,''
{\it JHEP} \textbf{12}, 057 (2003).
%doi:10.1088/1126-6708/2003/12/057


%\cite{Han:2003wu}
\bibitem{Han:2003wu}
T.~Han, H.~E.~Logan, B.~McElrath and L.~T.~Wang,
%``Phenomenology of the little Higgs model,''
{\it Phys. Rev. } D \textbf{67}, 095004 (2003).
%doi:10.1103/PhysRevD.67.095004


%\cite{Chang:2003un}
\bibitem{Chang:2003un}
S.~Chang and J.~G.~Wacker,
%``Little Higgs and custodial SU(2),''
{\it Phys. Rev.}  D \textbf{69}, 035002 (2004).
%doi:10.1103/PhysRevD.69.035002


%\cite{Schmaltz:2004de}
\bibitem{Schmaltz:2004de}
M.~Schmaltz,
%``The Simplest little Higgs,''
{\it JHEP} \textbf{08}, 056 (2004).
%doi:10.1088/1126-6708/2004/08/056


\bibitem{Schmaltz:2008vd}
M.~Schmaltz and J.~Thaler,
%``Collective Quartics and Dangerous Singlets in Little Higgs,''
{\it JHEP} \textbf{03}, 137 (2009).
%doi:10.1088/1126-6708/2009/03/137



\bibitem{JHEP09-2010} M.~Schmaltz, D.~Stolarski and J.~Thaler,
%``The Bestest Little Higgs,''
{\it JHEP} \textbf{09}, 018 (2010).
%doi:10.1007/JHEP09(2010)018


%\cite{Godfrey:2012tf}
\bibitem{Godfrey:2012tf}
S.~Godfrey, T.~Gregoire, P.~Kalyniak, T.~A.~W.~Martin and K.~Moats,
%``Exploring the heavy quark sector of the Bestest Little Higgs model at the LHC,''
{\it JHEP} \textbf{04}, 032 (2012).
%doi:10.1007/JHEP04(2012)032


\bibitem{Kalyniak:2013eva} P.~Kalyniak, T.~Martin and K.~Moats,
%``Constraining the Little Higgs model of Schmaltz, Stolarski, and Thaler with recent results from the LHC,''
{\it Phys. Rev.} \textbf{D} 91, 013010 (2015).
%doi:10.1103/PhysRevD.91.013010

%\cite{Martinez-Martinez:2024lez}
\bibitem{Martinez-Martinez:2024lez}
J.~M.~Mart\'\i{}nez-Mart\'\i{}nez, A.~Guti\'errez-Rodr\'\i{}guez, E.~Cruz-Albaro and M.~A.~Hern\'andez-Ru\'\i{}z,
%``New Z' boson of the bestest little Higgs model as a portal to signatures of Higgs bosons h $_{0}$ and H $_{0}$ at the future muon collider,''
{\it Chin. Phys.} C \textbf{49},  073101 (2025).
%doi:10.1088/1674-1137/adcc8a

%\cite{Cruz-Albaro:2024vjk}
\bibitem{Cruz-Albaro:2024vjk}
E.~Cruz-Albaro, A.~Guti\'errez-Rodr\'\i{}guez, D.~Espinosa-G\'omez, T.~Cisneros-P\'erez and F.~Ram\'\i{}rez-Zavaleta,
%``Probing heavy Higgs boson production and decay in the bestest little Higgs model at the LHC,''
{\it Phys. Rev.} D \textbf{110},  015013 (2024).
%doi:10.1103/PhysRevD.110.015013

%%\cite{Cisneros-Perez:2024efk}
%\bibitem{Cisneros-Perez:2024efk}
%T.~Cisneros-P\'erez, E.~Cruz-Albaro, A.~Y.~Ojeda-Casta\~neda and S.~E.~Sol\'\i{}s-N\'u\~nez,
%%``Chromomagnetic dipole moments of light quarks in the bestest little Higgs model,''
%{\it Chin. Phys.} C \textbf{48},  103109 (2024).
%%doi:10.1088/1674-1137/ad62d9
%
%%\cite{Cisneros-Perez:2023foe}
%\bibitem{Cisneros-Perez:2023foe}
%T.~Cisneros-P\'erez, M.~A.~Hern\'andez-Ru\'\i{}z, A.~Guti\'errez-Rodr\'\i{}guez and E.~Cruz-Albaro,
%%``Flavor-changing top quark rare decays in the Bestest Little Higgs Model,''
%{\it Eur. Phys. J.} C \textbf{83},  1093 (2023).
%%doi:10.1140/epjc/s10052-023-12264-6
%
%
%
%\cite{Cruz-Albaro:2022lks}
\bibitem{Cruz-Albaro:2022lks}
E.~Cruz-Albaro, A.~Guti\'errez-Rodr\'\i{}guez, J.~I.~Aranda and F.~Ram\'\i{}rez-Zavaleta,
%``Research on the electromagnetic and weak dipole moments of the tau-lepton at the Bestest Little Higgs Model,''
{\it Eur. Phys. J.} C \textbf{82}, 1095 (2022).
%doi:10.1140/epjc/s10052-022-11076-4




%\cite{ZurbanoFernandez:2020cco}
\bibitem{ZurbanoFernandez:2020cco}
I.~Zurbano Fernandez, M.~Zobov, A.~Zlobin, F.~Zimmermann, M.~Zerlauth, C.~Zanoni, C.~Zannini, O.~Zagorodnova, I.~Zacharov and M.~Yu, \textit{et al.},
{\it High-Luminosity Large Hadron Collider (HL-LHC): Technical design report},
CERN, 2020,
ISBN 978-92-9083-586-8, 978-92-9083-587-5.
%doi:10.23731/CYRM-2020-0010



%\cite{Azzi:2019yne}
\bibitem{Azzi:2019yne}
P.~Azzi, S.~Farry, P.~Nason, A.~Tricoli, D.~Zeppenfeld, R.~Abdul Khalek, J.~Alimena, N.~Andari, L.~Aperio Bella and A.~J.~Armbruster, \textit{et al.},
%``Report from Working Group 1: Standard Model Physics at the HL-LHC and HE-LHC,''
{\it CERN Yellow Rep. Monogr.} \textbf{7}, 1 (2019).
%doi:10.23731/CYRM-2019-007.

%\cite{Cepeda:2019klc}
\bibitem{Cepeda:2019klc}
M.~Cepeda, S.~Gori, P.~Ilten, M.~Kado, F.~Riva, R.~Abdul Khalek, A.~Aboubrahim, J.~Alimena, S.~Alioli and A.~Alves, \textit{et al.},
%``Report from Working Group 2: Higgs Physics at the HL-LHC and HE-LHC,''
{\it CERN Yellow Rep. Monogr. } \textbf{7}, 221 (2019).
%doi:10.23731/CYRM-2019-007.221


%\cite{FCC:2018bvk}
\bibitem{FCC:2018bvk}
A.~Abada \textit{et al.} (FCC Collaboration),
%``HE-LHC: The High-Energy Large Hadron Collider: Future Circular Collider Conceptual Design Report Volume 4,''
{\it Eur. Phys. J. ST} \textbf{228},  1109 (2019).
%doi:10.1140/epjst/e2019-900088-6






%\cite{MammenAbraham:2024gun}
\bibitem{MammenAbraham:2024gun}
R.~Mammen Abraham, J.~Adhikary, J.~L.~Feng, M.~Fieg, F.~Kling, J.~Li, J.~Pei, T.~R.~Rabemananjara, J.~Rojo and S.~Trojanowski,
%``FPF@FCC: neutrino, QCD, and BSM physics opportunities with far-forward experiments at a 100 TeV Proton Collider,''
{\it JHEP} \textbf{01}, 094 (2025).
%doi:10.1007/JHEP01(2025)094

%\cite{Mangano:2022ukr}
\bibitem{Mangano:2022ukr}
M.~L.~Mangano, W.~Riegler, M.~Aleksa, P.~P.~Allport, S.~Asai, A.~Ball, M.~I.~Besana, E.~R.~Bielert, S.~Bologna and E.~Boos, \textit{et al.},
{\it Conceptual design of an experiment at the FCC-hh, a future 100 TeV hadron collider},
doi:10.23731/CYRM-2022-002.


%\cite{FCC:2018byv}
\bibitem{FCC:2018byv}
A.~Abada \textit{et al.} (FCC Collaboration),
%``FCC Physics Opportunities: Future Circular Collider Conceptual Design Report Volume 1,''
{\it Eur. Phys. J.} C \textbf{79}, 474 (2019).
%doi:10.1140/epjc/s10052-019-6904-3


%\cite{FCC:2018vvp}
\bibitem{FCC:2018vvp}
A.~Abada \textit{et al.} (FCC Collaboration),
%``FCC-hh: The Hadron Collider: Future Circular Collider Conceptual Design Report Volume 3,''
{\it Eur. Phys. J. ST } \textbf{228},  755 (2019).
%doi:10.1140/epjst/e2019-900087-0




%\cite{ATLAS:2012ad}
\bibitem{ATLAS:2012ad}
G.~Aad \textit{et al.} (ATLAS Collaboration),
%``Search for the Standard Model Higgs boson in the diphoton decay channel with 4.9 fb$^{-1}$ of $pp$ collisions at $\sqrt{s}=7$ TeV with ATLAS,''
{\it Phys. Rev. Lett.} \textbf{108}, 111803 (2012).
%doi:10.1103/PhysRevLett.108.111803



%\cite{ATLAS:2015yey}
\bibitem{ATLAS:2015yey}
G.~Aad \textit{et al.} (ATLAS and CMS Collaborations),
%``Combined Measurement of the Higgs Boson Mass in $pp$ Collisions at $\sqrt{s}=7$ and 8 TeV with the ATLAS and CMS Experiments,''
{\it Phys. Rev. Lett.} \textbf{114}, 191803 (2015).
%doi:10.1103/PhysRevLett.114.191803

%\cite{Arhrib:2018pdi}
\bibitem{Arhrib:2018pdi}
A.~Arhrib, R.~Benbrik, J.~El Falaki, M.~Sampaio and R.~Santos,
%``Pseudoscalar decays to gauge bosons at the LHC and at a future 100 TeV collider,''
{\it Phys. Rev.} D \textbf{99},  035043 (2019).
%doi:10.1103/PhysRevD.99.035043

%\cite{Abu-Ajamieh:2025zcv}
\bibitem{Abu-Ajamieh:2025zcv}
F.~Abu-Ajamieh, S.~Modak, S.~Mukherjee and S.~K.~Vempati,
{\it Pseudoscalar Higgs Production at Muon Colliders: The Role of One-Loop Effective Vertices},
arXiv:2505.02092 [hep-ph].

%\cite{Aranda:2017bgq}
\bibitem{Aranda:2017bgq}
J.~I.~Aranda, E.~Cruz-Albaro, D.~Espinosa-G\'omez, J.~Monta\~no, F.~Ram\'\i{}rez-Zavaleta and E.~S.~Tututi,
%``Heavy neutral pseudoscalar decays into gauge bosons in the Littlest Higgs model,''
{\it J. Phys. } G \textbf{44},  105002 (2017).
%doi:10.1088/1361-6471/aa846d

%\cite{Cornell:2020usb}
\bibitem{Cornell:2020usb}
A.~S.~Cornell, A.~Deandrea, B.~Fuks and L.~Mason,
%``Future lepton collider prospects for a ubiquitous composite pseudoscalar,''
{\it Phys. Rev.} D \textbf{102},  035030 (2020).
%doi:10.1103/PhysRevD.102.035030


%\cite{Almarashi:2021pgu}
\bibitem{Almarashi:2021pgu}
M.~M.~Almarashi,
%``Review of a Light NMSSM Pseudoscalar Higgs-State Production at the LHC,''
{\it Universe} \textbf{7}, 392 (2021).
%doi:10.3390/universe7110392


%\cite{ATLAS:2024jja}
\bibitem{ATLAS:2024jja}
G.~Aad \textit{et al.} (ATLAS Collaboration),
%``Search for $t\bar{t}H/A \rightarrow t\bar{t}t\bar{t}$ production in proton{\textendash}proton collisions at $\sqrt{s}=13$~$\text {TeV}$ with the ATLAS detector,''
{\it Eur. Phys. J.} C \textbf{85},  573 (2025).
%doi:10.1140/epjc/s10052-025-14041-z


%\cite{ATLAS:2022rws}
\bibitem{ATLAS:2022rws}
G.~Aad \textit{et al.} (ATLAS Collaboration),
%``Search for $ t\overline{t}H/A\to t\overline{t}t\overline{t} $ production in the multilepton final state in proton{\textendash}proton collisions at $ \sqrt{s} $ = 13 TeV with the ATLAS detector,''
{\it JHEP } \textbf{07}, 203 (2023).
%doi:10.1007/JHEP07(2023)203

%\cite{ATLAS:2017snw}
\bibitem{ATLAS:2017snw}
M.~Aaboud \textit{et al.} (ATLAS Collaboration),
%``Search for Heavy Higgs Bosons $A/H$ Decaying to a Top Quark Pair in $pp$ Collisions at $\sqrt{s}=8\text{ }\text{ }\mathrm{TeV}$ with the ATLAS Detector,''
{\it Phys. Rev. Lett. } \textbf{119},  191803 (2017).
%doi:10.1103/PhysRevLett.119.191803

%\cite{Kao:2003jw}
\bibitem{Kao:2003jw}
C.~Kao, G.~Lovelace and L.~H.~Orr,
%``Detecting a Higgs pseudoscalar with a $Z$ boson at the LHC,''
{\it Phys. Lett.} B \textbf{567}, 259 (2003).
%doi:10.1016/j.physletb.2003.06.042



%\cite{Cruz-Albaro:2023pah}
\bibitem{Cruz-Albaro:2023pah}
E.~Cruz-Albaro, A.~Gutierrez-Rodr\i{}guez, M.~A.~Hernandez-Ru\i{}z and T.~Cisneros-Perez,
%``Searching the anomalous electromagnetic and weak dipole moments of the top quark at the bestest little Higgs model,''
{\it Eur. Phys. J. Plus} \textbf{138}, 506 (2023).
%doi:10.1140/epjp/s13360-023-04125-8



%\cite{Cruz-Albaro:2022kty}
\bibitem{Cruz-Albaro:2022kty}
E.~Cruz-Albaro and A.~Guti\'errez-Rodr\'\i{}guez,
%``Weak dipole moments of the top-quark at the Bestest Little Higgs model,''
{\it Eur. Phys. J. Plus } \textbf{137},  1295 (2022).
%doi:10.1140/epjp/s13360-022-03496-8


%\cite{Gutierrez-Rodriguez:2023sxg}
\bibitem{Gutierrez-Rodriguez:2023sxg}
A.~Guti\'errez-Rodr\'\i{}guez, E.~Cruz-Albaro, D.~Espinosa-G\'omez, T.~Cisneros-P\'erez and D.~A.~P\'erez-Carlos,
{\it New physics search with the new gauge boson $Z'$ of the bestest little Higgs model at the muon collider},
arXiv:2312.08560 [hep-ph].

\bibitem{pdg:2023}
R.~L.~Workman \textit{et al.} (Particle Data Group),
%``Review of Particle Physics,''
{\it Prog. Theor. Exp. Phys.} \textbf{2022}, 083C01 (2022).
%doi:10.1093/ptep/ptac097

%\cite{Barger:1987nn}
\bibitem{Barger:1996}
V.~D.~Barger and R.~J.~N.~Phillips,
{\it Collider Physics},
Addison-Wesley (1996); V.~D.~Barger and R.~J.~N.~Phillips,
{\it Collider Physics},
Addison-Wesley (1997).

%\cite{Barradas:1996xb}
\bibitem{Barradas:1996xb}
E.~Barradas, J.~L.~Diaz-Cruz, A.~Gutierrez and A.~Rosado,
%``Three body decays of Higgs bosons in the MSSM,''
{\it Phys. Rev.} D \textbf{53}, 1678 (1996).
%doi:10.1103/PhysRevD.53.1678

%\cite{Denner:2005nn}
\bibitem{Denner:2005nn}
A.~Denner and S.~Dittmaier,
%``Reduction schemes for one-loop tensor integrals,''
{\it Nucl. Phys.} B \textbf{734}, 62 (2006).
%doi:10.1016/j.nuclphysb.2005.11.007

%\cite{Patel:2015tea}
\bibitem{Patel:2015tea}
H.~H.~Patel,
%``Package-X: A Mathematica package for the analytic calculation of one-loop integrals,''
{\it Comput. Phys. Commun.} \textbf{197}, 276 (2015).
%doi:10.1016/j.cpc.2015.08.017

 %\cite{Patel:2016fam}
\bibitem{Patel:2016fam}
H.~H.~Patel,
%``Package-X 2.0: A Mathematica package for the analytic calculation of one-loop integrals,''
{\it Comput. Phys. Commun.} \textbf{218}, 66 (2017).
%doi:10.1016/j.cpc.2017.04.015


\bibitem{ATLAS:2020gxx}
G.~Aad, \textit{et al.} (ATLAS Collaboration),
%``Search for a heavy Higgs boson decaying into a Z boson and another heavy Higgs boson in the $\ell \ell bb$ and $\ell \ell WW$ final states in $pp$ collisions at $\sqrt{s}=13$ $\text {TeV}$ with the ATLAS detector,''
{\it Eur. Phys. J.} \textbf{C81}, 396 (2021).
%doi:10.1140/epjc/s10052-021-09117-5



%\cite{CMS:2019ogx}
\bibitem{CMS:2019ogx}
A.~M.~Sirunyan \textit{et al.} (CMS Collaboration),
%``Search for new neutral Higgs bosons through the H$\to$ ZA $\to \ell^{+}\ell^{-} \mathrm{b\bar{b}}$ process in pp collisions at $\sqrt{s} =$ 13 TeV,''
{\it JHEP} \textbf{03}, 055 (2020).
%doi:10.1007/JHEP03(2020)055




\end{thebibliography}
\end{document}